\documentclass[10pt]{article}         
\usepackage[utf8]{inputenc}
\usepackage{pifont}
\usepackage{amssymb}
\usepackage{amsmath}
\usepackage[english]{babel}
\usepackage[dvips]{graphicx}

\usepackage{fancyhdr}
\usepackage{titlesec}
\usepackage{vmargin}

\usepackage{subfig}

\usepackage{multirow}
\usepackage{flafter}
\usepackage{eurosym} 
\usepackage{array}

\usepackage{color}
\usepackage[%
    colorlinks=true,
    linkcolor=blue,
    citecolor=magenta]{hyperref}

\usepackage{pstricks}

\usepackage{psfrag}

\makeatletter
\renewcommand{\thesection}{\@arabic\c@section}
\makeatother

\DeclareFixedFont{\chapnumfont}{OT1}{phv}{b}{n}{80pt}
\DeclareFixedFont{\chapchapfont}{OT1}{phv}{m}{it}{40pt}
\DeclareFixedFont{\chaptitfont}{OT1}{phv}{b}{n}{25pt}

\definecolor{gris}{gray}{0.75}
\titleformat{\chapter}[display]%
	{\sffamily}%
	{\filleft\chapchapfont\color{gris}\chaptertitlename\
	\\
	\vspace{12pt}
	\chapnumfont\thechapter}%
	{16pt}%
	{\filleft\chaptitfont}%
	[\vspace{6pt}\titlerule\titlerule\titlerule]

\makeatletter
\renewcommand{\section}{\@startsection{section}{0}{0mm}%
{\baselineskip}{.5\baselineskip}%
{ \sffamily\Large\textbf}}%
\makeatother

\makeatletter
\renewcommand{\subsection}{\@startsection{subsection}{1}{2mm}%
{\baselineskip}{.3\baselineskip}%
{\sffamily\Large\textbf}}%
\makeatother

\makeatletter
\renewcommand{\subsubsection}{\@startsection{subsubsection}{2}{4mm}%
{\baselineskip}{.15\baselineskip}%
{\sffamily\large\textbf}}%
\makeatother

\makeatletter
\renewcommand{\paragraph}{\@startsection{paragraph}{3}{6mm}%
{\baselineskip}{.15\baselineskip}%
{\sffamily\large\textbf}}%
\makeatother

\setmarginsrb{2.5cm}{1.5cm}{2.5cm}{2cm}{1cm}{1cm}{1cm}{1cm}

\begin{document}

\begin{center}
\Large{%
 Kinematic modelling of a 3-axis NC machine tool in linear and
  circular interpolation} \\
\vspace{.5cm}
\large{The International Journal of Advanced Manufacturing Technology}\\
\vspace{.5cm}
Xavier Pessoles, Yann Landon and Walter Rubio\\
 Universit\'e de Toulouse; INSA, UPS, Mines Albi, ISAE; ICA (Institut Cl\'ement
Ader);  135, avenue de Rangueil, F-31077 Toulouse, France\\
              Tel.: +33-(0)5-61558176\\
              \url{pessoles@lgmt.ups-tlse.fr} 
\end{center}





\subsection*{Abstract}
Machining time is a major performance criterion when it comes to high speed machining. 
CAM software can help in estimating that time for a given strategy. 
But in practice, CAM programmed feed rates are rarely achieved, especially where complex 
surface finishing is concerned. This means that machining time forecasts are often more 
than one step removed from reality. The reason behind this is that CAM routines do not 
take either the dynamic performances of the machines or their specific machining tolerances into account. 
The present article seeks to improve simulation of high speed NC machine dynamic behaviour and machining 
time prediction, offering two models. The first contributes through enhanced simulation of 3-axis paths in 
linear and circular interpolation, taking high speed machine accelerations and jerks into account. 
The second model allows transition passages between blocks to be integrated in the simulation by adding 
in a polynomial transition path that caters for the true machining environment tolerances. 
Models are based on respect for path monitoring. Experimental validation shows the contribution of polynomial modelling 
of the transition passage due to the absence of a leap in acceleration.
Simulation error on the machining time prediction remains below 1\%.

\subsection*{Keywords}
High speed machining; Linear interpolation; Circular interpolation; Polynomial
transition

\subsection*{Nomenclature}
\begin{center}
\begin{tabular}{p{0.1\textwidth}p{0.35\textwidth}}   
  \multicolumn{2}{l}{\textit{Kinematic and dynamic parameters}}\\
\hline
  $\overrightarrow{J}$ & jerk vector \\
  $\overrightarrow{A}$ & acceleration vector \\
  $\overrightarrow{V}$ & feed rate vector \\
  $\overrightarrow{X}$ & position vector \\
  $J_{max,i}$ & maximum jerk limited by machine dynamics on the axis  $i$ \\
  $A_{max,i}$ & maximum acceleration limited by the machine dynamics on the axis  $i$\\
  $A_{0,i}$, $V_{0,i}$, $X_{0,i}$ & acceleration, feed rate and initial position on the axis $i$ \\ 
    $V_F$ & programmed feed rate \\
  $V_F'$ & feed rate reached if $V_F$ is not achieved \\
  $V_{c,i}$ & feed rate set on the axis $i$ \\
  $V_{In}$ & feed rate of entry into a block\\
  $V_{Out}$ & feed rate exiting a block\\
  $V_{Out}'$ & feed rate exiting a block if $V_{Out}$ is not reached \\
  $V_{disc}$ & maximum feed rate for entering a discontinuity \\
  $V_{j}$ & feed rate entering a discontinuity limited by jerk \\
  $V_{a}$ & feed rate entering a discontinuity limited by acceleration \\
  $\tau_i$ & duration of phase $i$ ($\tau_i = T_i - T_{i-1}$)\\
\hline
\end{tabular}
\end{center}

\begin{center}
 \begin{tabular}{p{0.1\textwidth}p{0.35\textwidth}}
  \multicolumn{2}{l}{\textit{Circular transitions in linear interpolation }}\\
\hline
  $A,O,B$ & theoretical programmed path\\
  $A,Q,B$ & path described by the machine\\
  $TIT$, $tol_x$, $tol_y$ & method to define point $Q$ in accordance with the programming method\\
  $R$ & radius of the arc inserted on transition\\
  $l_1$, $l_2$ & length covered before and after the transition \\
  $\beta$ & angle formed by segments $[AO]$ and $[OB]$\\
  $a_c$, $j_t$ & normal acceleration and tangential jerk in steady state \\ 
\hline
\end{tabular}
\end{center}

\begin{center}
\begin{tabular}{p{0.1\textwidth}p{0.35\textwidth}}
  \multicolumn{2}{l}{\textit{Circular transitions in circular interpolation }}\\
\hline
  $R_1$, $R_2$ & radii of circles before and after a circle - circle
  transition \\ 
\hline
\end{tabular}
\end{center}

\begin{center}
\begin{tabular}{p{0.1\textwidth}p{0.35\textwidth}}
  \multicolumn{2}{l}{\textit{Parameters used in circular interpolation}}\\
\hline
  $O$ & centre of circle \\
  $R$ & radius of circle \\
  $P(t)$ & current point \\
  $\alpha$ & angle covered by the circle arc \\
  $\theta(t)$ & current angle \\
  $J_i$ & curvilinear jerk limited by the axis $i$ \\
  $J_C$ & curvilinear jerk \\
  $A_i$ & curvilinear acceleration limited by the axis $i$ \\
  $A_C$ & curvilinear acceleration \\
\hline
\end{tabular}
\end{center}

\begin{center}
\begin{tabular}{p{0.1\textwidth}p{0.35\textwidth}}
  \multicolumn{2}{l}{\textit{Polynomial transitions in linear
      interpolation}}\\  
\hline
$\mathcal{R}$ &  frame 
  $\left(O,\overrightarrow{X},\overrightarrow{Y},\overrightarrow{Z}\right)$ \\
  $A,O,B$ & programmed theoretical path\\
  $(x_A,y_A,z_A)$ & coordinates of $\overrightarrow{OA}$ in the frame $\mathcal{R}$ \\
  $(x_B,y_B,z_B)$ & coordinates of $\overrightarrow{OB}$ in the frame $\mathcal{R}$ \\
  $M$ & point of entry into the discontinuity\\
  $N$ & point of exit from the discontinuity\\
  $Q$ & effective point of passage in the discontinuity\\
  $T$ & time of passage in the discontinuity\\
  $L$ & distance $OM$\\
  $P(t)$ & current point \\
  $\phi_i, \theta_i$ & spherical coordinates of the point $i$\\
  $tol_x, tol_y, tol_z$ & tolerance of position on the axes\\
  $V_M, V_N$ & feed rates for entry on M and exit on N\\
\hline
\end{tabular}
\end{center}

\section{Introduction}
\label{sec:int}
High speed machining centres allow for extremely high feed rates to be programmed. However, when machining molds or dies, dynamic performances of the machines do not always allow such feed rates to be reached. Indeed, according to the quality sought, the segments making up the machining path are often extremely short and in such conditions the feed rate reached by the machine will then be limited by the NC interpolation time or even the capabilities in jerk or acceleration of the axes 
\cite{tapie_circular_2007}, \cite{flores_evaluation_2007}. The feed rate will then not be constant, 
leading to considerably lower productivity, a
variation in tangential cutting forces and impaired quality
\cite{korkut_influence_2007}. Many publications relating to the search to reduce the number of feed rate changes base their research on the use of NURBS interpolations. 
\cite{tsai_real-time_2003} \cite{liu_adaptive_2005}
\cite{xu_adaptive_2008} or B-spline \cite{sencer_feed_2008}. However, in the industrial world, linear and circular interpolation remain the most frequently used methods on many workpieces. Precise simulation of this type of movement is therefore essential. 

The aim of this work is therefore to propose a comprehensive model intended to simulate the position, feed rate, acceleration and jerk in 3-axis linear and circular interpolation taking the machine/NC combination parameters into account.

\begin{figure}[!ht]
  \centering
  \caption{Control law for a translation axis}
  \includegraphics[width=.4\textwidth]{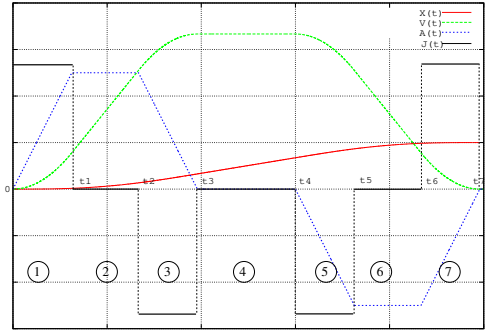}      
  \label{fig:loicommande}
\end{figure}

At present, NC machine manufacturers  \cite{www:siemens} propose a displacement law on the axes in trapezoid acceleration. 
This type of command has been studied in the literature by a number of authors writing on uniaxial paths with null initial and final feed rates \cite{erkorkmaz_high_2001} \cite{aguilar2006commande}.
This movement involves 7 phases (figure \ref{fig:loicommande}):

\begin{itemize}
\item On phase 1,
\begin{equation}
\forall t\in\left[T_0,T_1\right]
\left\{
  \begin{array}{lcl}
    J_i(t) & = & J_{max,i} \\
    A_i(t) & = & A_{0,i} + J_{max,i}\left(t -T_0\right)\\
    V_i(t) & = & V_{0,i} + A_{0,i}\left(t -T_0\right) \\
&&    +\frac{1}{2}J_{max,i}\left(t-T_0\right)^2\\ 
    X_i(t) & = & X_{0,i} + V_{0,i}\left(t -T_0\right) +\\
&& \frac{1}{2}A_{0,i}\left(t -T_0\right)^2+ \\
&& \frac{1}{6}J_{max,i}\left(t -T_0\right)^3\\
  \end{array}
\right.
\label{eq:phase1}
\end{equation}
\item On phase 2,
\begin{equation}
\forall t\in\left[T_1,T_2\right]
\left\{
  \begin{array}{lcl}
    J_i(t) & = & 0 \\
    A_i(t) & = & A_{max,i}\\
    V_i(t) & = & V_i\left(T_1\right) + A_{max,i}\left(t-T_1\right)\\
    X_i(t) & = & X_i\left(T_1\right) + V_i(T_1)\left(t-T_1\right)\\
&&    +\frac{1}{2} A_{max,i}\left(t-T_1\right)^2\\ 
  \end{array}
\right.
\label{eq:phase2}
\end{equation}
\item On phase 3, 
\begin{equation}
\forall t\in\left[T_2,T_3\right]
\left\{
  \begin{array}{lcl}
    J_i(t) & = & -J_{max,i} \\
    A_i(t) & = & A_i\left(T_2\right) -J_{max,i} \left(t-T_2\right)  \\
    V_i(t) & = & V_i\left(T_2\right) + A_i(T_2)\left(t-T_2\right)   \\
&&    -\frac{1}{2}J_{max,i} \left(t-T_2\right)^2  \\ 
    X_i(t) & = & X_i\left(T_2\right) +
    V_i\left(T_2\right)\left(t-T_2\right) + \\
&&    \frac{1}{2}A_i(T_2)\left(t-T_2\right)^2 \\
&&    -\frac{1}{6}J_{max,i} \left(t-T_2\right)^3  \\
  \end{array}
\right.
\label{eq:phase3}
\end{equation}
\item On phase 4, 
\begin{equation}
\forall t\in\left[T_3,T_4\right]
\left\{
  \begin{array}{lcl}
    J_i(t) & = & 0 \\
    A_i(t) & = & 0 \\
    V_i(t) & = & V_{c,i} \\
    X_i(t) & = & X_i\left(T_3\right) +  V_{c,i}\left(t-T_3\right) 
  \end{array}
\right.
\label{eq:phase4}
\end{equation}
\item On phase 5, 
\begin{equation}
\forall t\in\left[T_4,T_5\right]
\left\{
  \begin{array}{lcl}
    J_i(t) & = & -J_{max,i} \\
    A_i(t) & = & A_i \left(T_4 \right) - J_{max,i}\left(t -T_4\right)\\
    V_i(t) & = & V_i \left(T_4 \right) + 
    A_i \left(T_4 \right) \left(t -T_4\right) \\
&&    -\frac{1}{2}J_{max,i}\left(t -T_4\right)^2\\
    X_i(t) & = & X_i\left(T_4\right)+V_i\left(T_4\right)\left(t
      -T_4\right) \\
&&    +\frac{1}{2}A_i\left(T_4\right)\left(t-T_4\right)^2\\
&&    -\frac{1}{6}J_{max,i}\left(t -T_4\right)^3
  \end{array}
\right.
\label{eq:phase5}
\end{equation}
\item On phase 6,
\begin{equation}
\forall t\in\left[T_5,T_6\right]
\left\{
  \begin{array}{lcl}
    J_i(t) & = & 0 \\
    A_i(t) & = & -A_{max,i}\\
    V_i(t) & = & V_i\left(T_5\right) - A_{max,i}\left(t-T_5\right)\\
    X_i(t) & = & X_i\left(T_5\right) + V_i(T_5)\left(t-T_5\right)\\
&&    -\frac{1}{2} A_{max,i}\left(t-T_5\right)^2\\ 
  \end{array}
\right.
\label{eq:phase6}
\end{equation}
\item On phase 7,
\begin{equation}
\forall t\in\left[T_6,T_7\right]
\left\{
  \begin{array}{lcl}
    J_i(t) & = & J_{max,i} \\
    A_i(t) & = & A_i \left(T_6 \right) + J_{max,i}\left(t -T_6\right)\\
    V_i(t) & = & V_i \left(T_6 \right) + 
    A_i \left(T_6 \right) \left(t -T_6\right)   \\
&&    +\frac{1}{2}J_{max,i}\left(t -T_6\right)^2\\
    X_i(t) & = & X_i\left(T_6\right)+V_i\left(T_6\right)\left(t
      -T_6\right) \\
&&    +\frac{1}{2}A_i\left(T_6\right)\left(t-T_6\right)^2\\
&&    +\frac{1}{6}J_{max,i}\left(t -T_6\right)^3
  \end{array}
\right.
\label{eq:phase7}
\end{equation}
\end{itemize}

Acceleration starts by increasing up to its maximum value $A_{max,i}$. The slope for this acceleration is called jerk. This is noted $J_{max,i}$ (phase 1). An acceleration
 level is then reached with value $A_{max,i}$ (phase 2). This then diminishes down to zero (phase 3). During 
the fourth stage, acceleration is null and the feed rate attained is equal to the programmed feed rate. Phases 5, 6 and 7 are symmetrical with phases 3, 2 and 1. Time resolution of this system of equations allows the times for each of the phases to be calculated readily (section \ref{subsec:linear}). Parameters $A_{max,i}$ and $J_{max,i}$ are specific to the dynamics of the machine used.

In practice, the blocks follow on from each other and the feed rates at the start and end of the blocks are rarely null. Furthermore, according to the length of the segment and the feed rate to be reached, maximum acceleration or the programmed feed rate are not always reached. Thus, segments are often crossed in which some of the seven phases do not exist. This general case is rarely addressed \cite{erkorkmaz_high_2001}
\cite{nam_study_2004}. Modelling of the machine's behaviour taking these aspects into account is proposed in section \ref{sec:mod}. 

In the case of 3-axis machining, the path of a segment is
covered simultaneously on 3 axes. Take, for example, the case of a block with displacement on $\overrightarrow{X}$ and $\overrightarrow{Y}$. To ensure the quality of the machined workpiece, the path actually followed needs to be monitored in relation to the programmed path. To do so, synchronization of the axes is needed. Indeed, both axes must reach the final position at the same instant. 
 Thus, on 3 axes, a feed rate, an acceleration and a maximum jerk have to be calculated for each of the axes as a function of the slowest axis \cite{lavernhe_kinematical_2008}. Using the model in 7 phases, the same time for each of the phases will be obtained on each of the axes, allowing the path to be followed. 

Circular interpolation is also widely used. However, as far as can be ascertained, there are no publications covering changes in feed rate on a circle while also ensuring monitoring of the path to be followed. Modelling of such a case will be presented in section \ref{sec:mod}.
 
\begin{figure}[!ht]
  \centering
  \caption{Model of circular pass for a discontinuity at a tangent}
  \includegraphics[width=.4\textwidth]{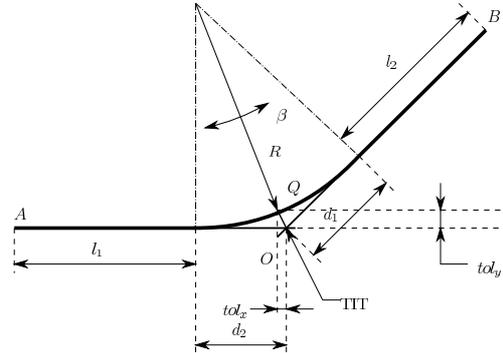}
  \label{fig:dugas}
\end{figure}

Moreover, a machining operation can be broken down into a multitude of linear or circular blocks. This thus poses the problem of the crossing of transitions between discontinuous blocks tangent to each other. From a purely kinematic point of view, the exact passage by the programmed points requires precise arrest of the machine at the end of each block. This behaviour is not permitted in practice as it implies repeatedly slowing down and thus a loss in productivity. In addition, the jerks are prejudicial to the quality of the workpieces manufactured as well as the lifetime of the cutting tools used. Thus, in numerical commands, there is a tolerance on pursuit of the path, figure \ref{fig:dugas}. For a programmed path $A$ -- $O$ --
$B$, the machine will cover the path $A$ -- $Q$ -- $B$.
This point $Q$ can be defined in a number of ways: by the distance $TIT$, 
tolerances $tol_x$ and $tol_y$ on the axes and the distances $d_1$ or
$d_2$. 

In the literature, a circle arc is used to model this transition \cite{manuel_influence_2003}. Radius $R$ of this circle arc is calculated in relation to the tolerance permitted by the NC and lengths $l_1$ and $l_2$ of the blocks before and after the discontinuity: 

\begin{equation}
R = \min\left(
  TIT\frac{\cos\left(\frac{\beta}{2}\right)}
  {1-\cos\left(\frac{\beta}{2}\right)}, 
  \frac{l}{\sin\left(\frac{\beta}{2}\right)}-TIT
\right)
\end{equation}
where $l=\min\left(l_1,l_2\right)$.

From a kinematic point of view, the course of a circle at constant feed rate $V$ runs at constant centripetal acceleration $a_c =
\frac{V_F^2}{R}$. Deriving this acceleration, a constant tangential jerk $j_t = \frac{V^3}{R^2}$ is also obtained. As this acceleration $a_c$ and jerk $j_t$ are limited on the machine by maximum values $A_{max,i}$ and $J_{max,i}$, for a given radius, the feed rate in the discontinuity will thus be limited: 

\begin{equation}
V_{disc}=\min \left(V_F, V_a = \sqrt{A_{max,i}R}, V_j=\sqrt[3]{J_{max,i}R^2}
\right)
\end{equation}

This modelling allows the maximum speed of passage to be expressed
as a discontinuity. However, it pre-supposes a jump in acceleration: at
constant feed rate on segment $[AO]$ acceleration is null (phase 4)
while at constant speed on a circle, the projection of the
acceleration vector on the axes can reach $V_F^2/R$. This leap in
acceleration on crossing the transition is not observed in
practice. Thus, a circular model cannot be used to simulate precisely
the position, feed rate, acceleration and jerk along the
discontinuity, even if it gives a good approximation of the drop in
feed rate. In part \ref{sec:discontpoly}, a polynomial form of modelling for transitions between non-tangent segments will be presented. 

In circular interpolation, the transitions between the circles of different radii need to be modelled. In this case, the same type of limitation arises: two tangent circles with different radii are discontinuous in curvature. On crossing the discontinuity, there will thus be a jump in acceleration. Pateloup \cite{pateloup_corner_2004} proposes a model to determine the minimum feed rate needed to cross the discontinuity taking into account the jerk $j$, the radii $R_1$ and $R_2$ of the two circular portions, and the interpolation time $\delta_t$ of the machine:

\begin{equation}
V_{disc} = \sqrt{\frac{R_1 R_2}{\left|R_1 -R_2\right|\delta_t j}}
\label{eq:patteloup}
\end{equation}

This model is also used to cross a transition between a segment of a straight line and a circle arc when they are tangent. The model is again taken up in the algorithm proposed by Tapie \cite{tapie_circular_2007} to calculate the entry rate into this type of discontinuity. This method for crossing discontinuities will be validated in section \ref{sec:validation}. 

With the aim of simulating a complete path (path including blocks and transitions), Lavernhe \cite{lavernhe_kinematical_2008} proposes a method allowing the machine's dynamic behaviour to be computed using a formalism with inverse time. Integration of the NC cycle time in this method allows it to predict the control jerk value to be predicted for each of the periods. From this are deduced the plots for acceleration and feed rate. Furthermore, his model takes the predictive functions available on NCs into account. The model for crossing of discontinuities in tangency is that described previously (circle arc). 

Another solution is to identify the servo-system model for the machine/NC combination \cite{tounsi_identification_2003}. This approach appears difficult to implement given the lack of data provided by NC manufacturers. Indeed, to apply this approach would require precise knowledge of the slaving flow diagrams for the axes and especially the various correctors used. Where appropriate, tests need to be conducted to identify the transfer function parameters.

Finally, a third method involves modelling directly the laws described in the previous sections. 
The difficulty in implementing these models lies in calculating for each block the time for each acceleration phase as 
well as the jerk on each axis. Integration of anticipation is no easy matter.

To sum up, the paths of linear blocks are clearly described in the
literature. However, the general case (path of a segment at non-null
initial and final feed rates) is not studied. Furthermore, no
information is to be found on the path of circular blocks. With
respect to discontinuities in tangency, the model for passage in a
circle arc allows the feed rate on crossing the discontinuity to be
quantified but does not enable laws for feed rates and accelerations
to be modelled. The idea is to propose algorithms that, on 3 axes,
show how to go from a feed rate on block entry $V_{In}$ to a feed rate
on block exit $V_{Out}$ while attempting to reach the programmed rate
$V_F$ both in linear interpolation (section \ref{subsec:linear}) and
circular interpolation (section \ref{subsec:circular}). A model is
then proposed for passage into discontinuities in tangency between two straight lines (section \ref{sec:discontpoly}). Finally, tests validating the simulator are presented.

\section{Modelling NC behaviour in linear and circular interpolation }
\label{sec:mod} 
In this section, the laws for displacements, feed rates, accelerations and jerks to cover a uniaxial segment from a feed rate $V_{In}$ to a feed rate $V_{Out}$ passing through a feed rate $V_F$ are modelled. Passage in the case of a 3-axis segment is then studied. The results are then adapted to circular interpolation.

\subsection{Modelling uniaxial linear segments in the general case}
\label{subsec:linear}

\begin{figure*}[!ht]
  \centering
  \caption{Set of cases of resolution}
  \includegraphics[width=.9\textwidth]{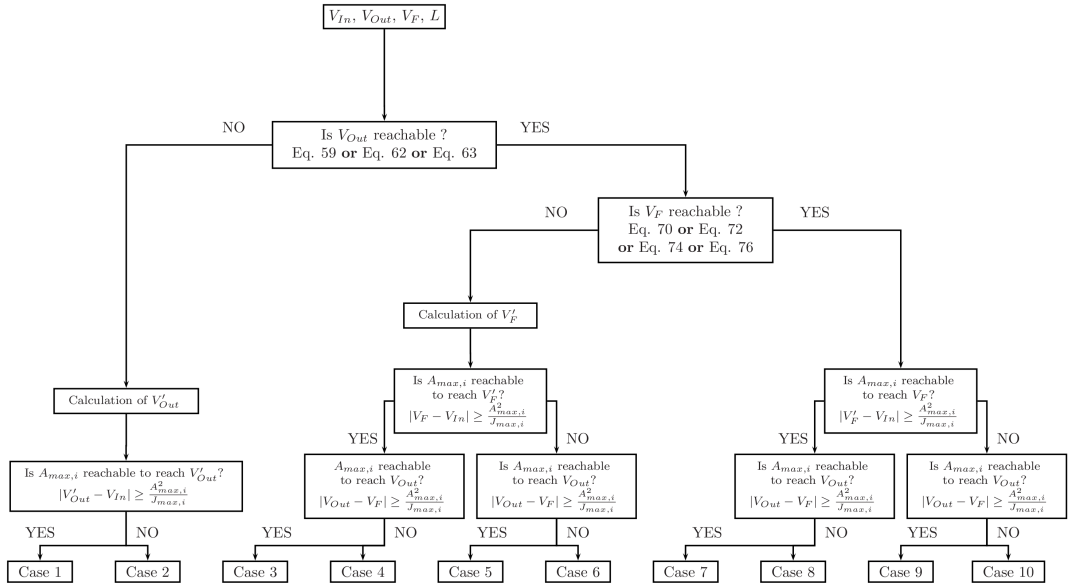}
  \label{fig:algocas}
\end{figure*}

In the general case of a toolpath for a segment in 3-axis machining, the feed rates at entry, middle and exit of a segment will not be identical. In addition, according to the jumps in feed rate to be crossed and the length of the displacement to be made, the feed rate will not necessarily be reached. As a result, resolution of equations \ref{eq:phase1} to \ref{eq:phase7} allows all existing cases to be identified (figure \ref{fig:algocas} and appendices). Thus, for $L$, $V_F$, $V_{In}$ and $V_{Out}$ given, it can be determined whether $V_{out}$ and/or $V_F$ will be reached going through $A_{max,i}$ or not. This means the duration for each of the phases can be known. The contour for jerk, acceleration, feed rate and position on one axis can thus be retraced.  

\subsection{Passage to 3 axes}
The algorithm given in the previous section is valid on a uniaxial displacement. 
In this case, values $A_{max,i}$ and $J_{max,i}$ will derive from the machine characteristics. 
On 3 axes, synchronization will be needed. This means that the times for each of the 7 phases are identical on all axes. 
As a result, for a given displacement at a given programmed feed rate, the NC will recalculate a set feed rate, acceleration and jerk for each of the axes. 

The modelling method followed is thus as follows: for a given segment, the displacement to be made on each of the axes is calculated. 
Using the results of the previous section with maximum acceleration and maximum jerk, it will thus be possible to determine which of 
the 3 axes will be the slowest. Then, using the results Lavernhe offers \cite{lavernhe_kinematical_2008}, feed rates $V_{i}$, accelerations $A_i$ and jerks $J_i$ on the axes limited $i$ can thus be calculated as a function of the distance $L_{lim}$ to be covered on the limiting axis and distances $L_i$ to be covered on the limited axes: 

\begin{equation}
V_{i} = \frac{L_i}{L_{lim}} V_{F}    \quad  
A_{i} = \frac{L_i}{L_{lim}} A_{max,i} \quad 
J_{i} = \frac{L_i}{L_{lim}} J_{max,i} 
\end{equation}
All the elements used to simulate tool paths in 3-axis linear interpolation have thus been presented. 
In what follows, the case of circular interpolation will be studied.

\subsection{Modelling tool paths defined by circle arcs }
\label{subsec:circular} 
As has been seen, few data are given as to simulation of displacements on a circle. The problem involves understanding how the axes of the NC behave to follow a circular path and especially how decelerations and accelerations are made when following a circle. To this purpose, the laws governing a circular movement can be stated. 

\begin{figure}[!ht]
  \centering
  \caption{Parametrization for a circular path}
  \includegraphics[width=.4\textwidth]{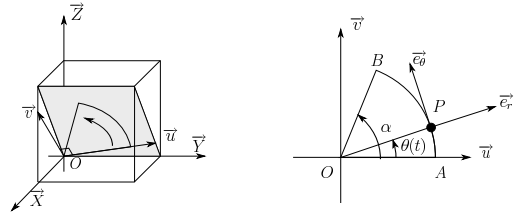}
  \label{fig:cercle}
\end{figure}

Let $P$ be a point moving over a circle arc of angle $\theta$, centre $O$ and radius $R$ in the frame $(\overrightarrow{e_r},\overrightarrow{e_{\alpha}},\overrightarrow{e_z})$
(figure \ref{fig:cercle}).

This gives:
\begin{equation}
\overrightarrow{OP} = R \overrightarrow{e_r}
\end{equation}

Note $\theta$ the angular law such that $\forall t \in [0,T]$, 
$\theta(0)=0$ and $\theta(T)=\alpha$. The feed rate, acceleration and jerk vectors can thus be expressed as follows:

\begin{eqnarray}
  \frac{d\overrightarrow{OP}}{dt} & = & 
  R \frac{d\theta}{dt} \overrightarrow{e_{\theta}} \label{eq:circle:v}\\
  \frac{d^2\overrightarrow{OP}}{dt^2} & = &
  R \frac{d^2\theta}{dt^2} \overrightarrow{e_{\theta}} - R
  \left(\frac{d\theta}{dt}\right)^2 
  \overrightarrow{e_{r}} \label{eq:circle:a}\\
  \frac{d^3\overrightarrow{OP}}{dt^3} & = & 
  R \left(
    \frac{d^3\theta}{dt^3} -\left(\frac{d\theta}{dt}\right)^3
  \right)
  \overrightarrow{e_{\theta}} 
  - 3R  \frac{d^2\theta}{dt^2}\frac{d\theta}{dt}
  \overrightarrow{e_{r}} \label{eq:circle:j}
\end{eqnarray}

When the set feed rate is reached, $d\theta (t)/dt = V_F /
R$. Nevertheless, the angle law $\theta (t)$ is unknown on passage from null feed rate to the programmed feed rate. As a result, using a law in 7 phases is proposed, like that considered in section \ref{subsec:linear}. In these conditions the jerk effect no longer corresponds to a jerk by axis, but a ``curvilinear'' jerk taking into account the influence of 2 or 3 axes along the plane in which the circle is made. This jerk is thus the result of several contributing axes. This means the tool path has to be expressed as a projection on the machine's axes of translation. This is done through two moves in frames.

 Let $\mathcal{M}_{1}$ and  $\mathcal{M}_{2}$ be the respective matrices for passage of the frame 
 $\left(\overrightarrow{u} ,\overrightarrow{v} ,\overrightarrow{w}
 \right)$ towards frame $\left(\overrightarrow{x} ,\overrightarrow{y}
   ,\overrightarrow{z} \right)$ and the frame 
 $\left(\overrightarrow{e_R} ,\overrightarrow{e_{\theta}}
   ,\overrightarrow{e_z} 
 \right)$ 
towards frame
$\left(\overrightarrow{u} ,\overrightarrow{v} ,\overrightarrow{w}
\right)$.

$\mathcal{M}$ is noted as the matrix for passage of the frame 
$\left(\overrightarrow{x} ,\overrightarrow{y} ,\overrightarrow{z}
\right)$ to frame $\left(\overrightarrow{e_R}
  ,\overrightarrow{e_{\theta}} ,\overrightarrow{e_z} 
\right)$:

\begin{equation}
\mathcal{M} = \left(\mathcal{M}_{1} \mathcal{M}_{2}\right)^{-1}
\end{equation}

\subsection{Calculation of jerk and curvilinear acceleration}
Curvilinear jerk and curvilinear acceleration do not form part of the machine parameters, but they are the result of the contribution made by accelerations and jerks for each axis in movement. According to the zone in which the circle is completed, one or other of the axes will be limiting. 

These parameters are calculated at the start and the end of movement as that is where they are most significant. Indeed, for the path of a circle arc, one needs to switch from a null normal acceleration to a normal acceleration equivalent to $V_F^2/R$ at the start and at the end of the path, which would require an infinite jerk at the start or the end of the path. As this can only be considered on certain types of machining centres, the maximum jerk that can be reached at the start and end of the path in consideration of the characteristics of the axes is thus calculated. 

Consider a circle arc made in the frame $\left(\overrightarrow{u} ,\overrightarrow{v},
  \overrightarrow{w} \right)$ from an angle $\theta(0) = 0$ to an angle
$\theta(T) = \alpha$, T being the overall duration of the path. In this general case, the matrix $\mathcal{M}$ is a matrix for rotation of the orthonormed frame $\left(\overrightarrow{e_R}
  ,\overrightarrow{e_{\theta}}, \overrightarrow{e_z} \right)$  to the orthonormed frame $\left(\overrightarrow{x} ,\overrightarrow{y},
 \overrightarrow{z} \right)$. This matrix is thus orthogonal and can be inverted. Note $(u_x, u_y, u_z)$, $(v_x, v_y, v_z)$,
$(w_x, w_y, w_z)$ the coordinates of the vectors 
$\overrightarrow{u}$, $\overrightarrow{v}$, $\overrightarrow{w}$ in the frame $\left(\overrightarrow{x} ,\overrightarrow{y},
  \overrightarrow{z} \right)$.

Using a model in 7 phases for angular acceleration and assuming initial acceleration to be null, the following is obtained on the first phase, $\forall
t\in[0,T_1]$:

\begin{equation}
\label{eq:jerkcurv}
\left\{
    \begin{array}{l}
      \frac{d^3\theta}{dt^3}(t) = J_c \\
      \frac{d^2\theta}{dt^2}(t) = J_c t \\
      \frac{d\theta}{dt}(t) =  \dot{\theta}_{In} + \frac{1}{2}J_c t^2 \\ 
      \theta(t)  =  \dot{\theta}_{In} t
      +\frac{1}{6}J_{c}t^3\\ 
    \end{array}
\right.
\end{equation}

with $J_c$ the curvilinear jerk. Carrying over equation
\ref{eq:jerkcurv} into equation
\ref{eq:circle:j} on $t=0$ and projecting onto the machine axes, the following is obtained: 

\begin{equation}
\frac{d^3\overrightarrow{OP}}{dt^3}  =  
  R \left( J_c -\dot{\theta}_{In}^3 \right)
  \overrightarrow{e_{\theta}}
  =\left( 
    \begin{array}{c}
      R \left( J_c -\dot{\theta}_{In}^3 \right) v_x \\
      R \left( J_c -\dot{\theta}_{In}^3 \right) v_x \\
      R \left( J_c -\dot{\theta}_{In}^3 \right) v_x
    \end{array}
\right)_{\left(\overrightarrow{x} ,\overrightarrow{y},
  \overrightarrow{z} \right)}
\end{equation}

Jerk is limited on each of the axes and curvilinear jerk will depend on 2 or even 3 axes. The latter will thus depend on the axis that will be limiting. The following will therefore obtain:
\begin{equation}
J_1 = \frac{J_{max,x}}{R v_x} + \dot{\theta}_{In}^3
\quad
J_2 = \frac{J_{max,y}}{R v_y} + \dot{\theta}_{In}^3
\quad
J_3 = \frac{J_{max,z}}{R v_z} + \dot{\theta}_{In}^3
\end{equation}

Similarly, $J_4$, $J_5$, $J_6$ are calculated at the end of movement to obtain: 

\begin{equation}
J_c = \min\limits_{i\in[1,6]} \left( J_i \right)
\end{equation}

Curvilinear acceleration $A_c$ can now be calculated. Noting $\theta_1$, the position reached at the end of phase 1 and $\dot{\theta_1}$ the feed rate reached at the end of phase 1, the following will obtain in phase 2, $\forall t \in [T_1,T_2]$:

\begin{equation}
\left\{
  \begin{array}{l}
    \frac{d^3\theta}{dt^3}(t) = 0 \\
    \frac{d^2\theta}{dt^2}(t) = A_{c}\\
    \frac{d\theta}{dt}(t) =  \dot{\theta_1} + A_{c} t\\
    \theta(t) = \theta_1 + \dot{\theta_1} t  +\frac{1}{2} A_ct^2\\ 
  \end{array}
\right.
\label{eq:acccurv}
\end{equation}

Carrying over equation \ref{eq:acccurv} into equation \ref{eq:circle:a}, this gives $t=T_1$:  

\begin{equation}
  \begin{array}{rcl}
    \frac{d^2\overrightarrow{OP}}{dt^2} & = & R A_c
    \overrightarrow{e_{\theta}} - R \dot{\theta_1}^2 
    \overrightarrow{e_{r}} \\
    &=& 
      \Big(
        - \left( u_x \cos\theta_1 + v_x \sin\theta_1
        \right)R\dot{\theta_1^2} \\
        &&
        + \left( u_y \cos\theta_1 + v_y \sin\theta_1 \right)RA_c
      \Big)\cdot \overrightarrow{x} \\
      &&+\Big(
        - \left( -u_x \sin\theta_1 + v_x \cos\theta_1
        \right)R\dot{\theta_1^2}\\
&&
        + \left( -u_y \sin\theta_1 + v_y \cos\theta_1 \right)RA_c 
      \Big)\cdot \overrightarrow{y}\\
     && +\Big(
        - w_x R\dot{\theta_1^2} +w_y RA_c 
      \Big)\cdot \overrightarrow{z}\\
  \end{array}
\end{equation}
  
Thus: 

\begin{equation}
  \begin{array}{rcl}
    A_1 & = & \frac{A_{max,x} + \left( u_x \cos\theta + v_x \sin\theta
      \right)R\dot{\theta_1^2}}{\left( u_y \cos\theta + v_y \sin\theta
      \right)R} 
    \\
    A_2 & = & \frac{A_{max,y} + \left(
        -u_x \sin\theta_1 + v_x \cos\theta_1
      \right)R\dot{\theta_1^2}}{\left(
        -u_y \sin\theta_1 + v_y \cos\theta_1
      \right)R} 
    \\ 
    A_3 & = &
    \frac{A_{max,z} + w_x R\dot{\theta_1^2}}{w_y R}
  \end{array}
\end{equation}

The same method is adopted for the deceleration phase, giving: 

\begin{equation}
A_c = \min\limits_{i\in[1,6]} \left( A_i \right)
\end{equation}

To conclude, in the modelling proposed in circular interpolation, the angular 
law follows a movement in seven phases, with maximum jerk being the calculated value $J_c$ 
and maximum acceleration being the calculated value $A_c$. Knowing all 
the parameters of a circle arc (point of departure, point of arrival, radius, etc.), 
the machine's dynamic behaviour can be simulated throughout the circular path.

This part allows the laws of position, feed rate, acceleration and jerk on unique 
blocks in circular and linear interpolation to be simulated. What remains is to model 
the junctions between blocks. It has already been shown that the model for transition 
between two tangent paths functions. The following part of the article will cover how to model the passage between two linear segments.

\section{Modelling of the transition between two rectilinear blocks }
\label{sec:discontpoly}
In the literature, transitions between two segments are always modelled by circle arcs, 
though this does not seem to match the real behaviour of the machine. Knowing that NCs are capable of describing polynomials of degree 5, it is suggested that they be used to model discontinuities in tangency. This model should allow a criterion for feed rate for entry into the discontinuity to be defined as also the contour for position, feed rate, acceleration and jerk in the discontinuity.

\subsection{Notations and hypotheses}
\begin{figure}[!ht]
  \centering
  \caption{Parametrization of passage of discontinuity}
  \includegraphics[width=0.4\textwidth]{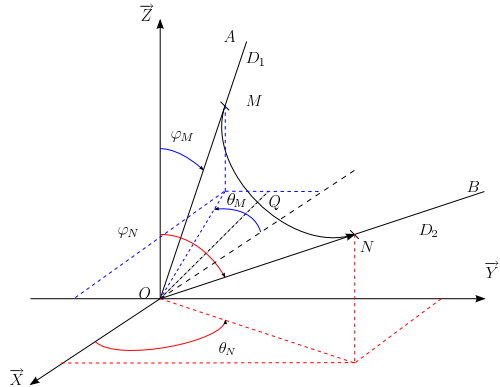}
  \label{fig:transition}
\end{figure}

Consider the paths of two consecutive non-aligned segments $AO$ and $OB$ (figure \ref{fig:transition}).

The points $M$ and $N$ can be expressed as follows in this frame:

\begin{equation}
  \begin{array}{c}
\overrightarrow{OM} = 
\left(
\begin{array}{c}
L\sin\varphi_M\cos\theta_M \\
L\sin\varphi_M\sin\theta_M \\
L\cos\varphi_M
\end{array}
\right)_{\mathcal{R}}
= L
\overrightarrow{u}
\\
\overrightarrow{ON} = 
\left(
\begin{array}{c}
L\sin\varphi_N\cos\theta_N \\
L\sin\varphi_N\sin\theta_N \\
L\cos\varphi_N
\end{array}
\right)_{\mathcal{R}}
= L \overrightarrow{v}
\end{array}
\end{equation}

These notations allow the working assumptions to be determined. Firstly, it shall be considered that: 
\begin{equation}
||\overrightarrow{OM}|| = ||\overrightarrow{ON}|| = L
\label{eq:hyp1}
\end{equation}

Then, considering the problem to be symmetrical, this gives:
\begin{equation}
\overrightarrow{OQ} =  ||\overrightarrow{OQ}|| \cdot 
\frac{\overrightarrow{u} + \overrightarrow{v}}{||\overrightarrow{u} +
  \overrightarrow{v}||} =  ||\overrightarrow{OQ}|| \cdot
\overrightarrow{w} = 
\left(
  \begin{array}{c}
    Q_x \\
    Q_y \\
    Q_z
  \end{array}
\right)_{\mathcal{R}}
\label{eq:hyp2}
\end{equation}

The coordinates of point $Q$ can then be expressed in the frame $\mathcal{R}$:

\begin{equation}
\begin{array}{c}
\overrightarrow{OQ} 
=
||\overrightarrow{OQ}|| \cdot
\left(
\begin{array}{c}
\sin\varphi_Q\cos \theta_Q \\
\sin\varphi_Q\sin \theta_Q \\
\cos\varphi_Q
\end{array}
\right)_{\mathcal{R}}
\quad
\\
\quad
\left\{
\begin{array}{c}
\varphi_Q  =  \arccos\frac{w_z}{ \sqrt{w_x^2+w_y^2+w_z^2}} \\
\mbox{Si } w_x \geq 0\, , \, \theta_Q =
\arcsin\frac{w_y}{\sqrt{w_x^2+w_y^2}} \\ 
\mbox{Si } w_x < 0\, , \, \theta_Q = \pi -
\arcsin\frac{w_y}{\sqrt{w_x^2+w_y^2}} \\ 
\end{array}
\right.
\end{array}
\end{equation}

The direction of the vector $\overrightarrow{OQ}$ is thus fully determined by the vectors $\overrightarrow{u}$ and $\overrightarrow{v}$. Its norm now needs to be determined. This is done by the tolerance granted the machine on passage of the discontinuity (section \ref{sec:int} and figure \ref{fig:loicommande}). In this instance, $tol_x$, $tol_y$ and $tol_z$ denote the maximum tolerances for passage used by the NC on each of the axes $\overrightarrow{X}$,
$\overrightarrow{Y}$ and 
$\overrightarrow{Z}$, which means that:

\begin{equation}
\left\{
\begin{array}{l}
\overrightarrow{OQ}\cdot\overrightarrow{X} \leq tol_x \\
\overrightarrow{OQ}\cdot\overrightarrow{Y} \leq tol_y \\
\overrightarrow{OQ}\cdot\overrightarrow{Z} \leq tol_z 
\end{array}
\right.
\end{equation}

The norm of vector $\overrightarrow{OQ}$ can thus be calculated as follows: 

\begin{equation}
||\overrightarrow{OQ}|| = \min\left( 
\frac{tol_x}{|\sin\varphi_Q \cos\theta_Q|}, 
\frac{tol_y}{|\sin\varphi_Q \sin\theta_Q|},
\frac{tol_z}{|\cos\varphi_Q|},
\right)
\end{equation}

Finally, due to the problem's symmetry, it can be considered that the entry feed rate in the discontinuity
$||\overrightarrow{V_M}||$ and the exit feed rate
$||\overrightarrow{V_N}||$ will be equal:  

\begin{equation}
\overrightarrow{V_M} = -V_{In}\overrightarrow{u} \quad
\overrightarrow{V_N} = V_{In}\overrightarrow{v}   
\label{eq:hyp3}
\end{equation}

As points $A$, $M$ and $O$, as well as points $B$, $N$ and $O$ are
aligned, angles $\varphi_M$, $\theta_M$, $\varphi_N$ and $\theta_N$ can thus be calculated:

\begin{equation}
\begin{array}{c}
\left\{
\begin{array}{c}
\varphi_M  =  \arccos\frac{z_A}{ \sqrt{x_A^2+y_A^2+z_A^2}} \\
\mbox{Si } x_A \geq 0\, , \, \theta_M = \arcsin\frac{y_A}{\sqrt{x_A^2+y_A^2}} \\
\mbox{Si } x_A < 0\, , \, \theta_M = \pi - \arcsin\frac{y_A}{\sqrt{x_A^2+y_A^2}} \\
\end{array}
\right.  
\\
\left\{
\begin{array}{c}
\varphi_N  =  \arccos\frac{z_B}{ \sqrt{x_B^2+y_B^2+z_B^2}} \\
\mbox{Si } x_B \geq 0\, , \, \theta_N = \arcsin\frac{y_B}{\sqrt{x_B^2+y_B^2}} \\
\mbox{Si } x_B < 0\, , \, \theta_N = \pi - \arcsin\frac{y_B}{\sqrt{x_B^2+y_B^2}} \\
\end{array}
\right.
\end{array}
\end{equation}

\subsection{Formation of the equation}

Using a polynomial representation, the equation for the position, feed
rate, acceleration and jerk of the point 
$P$ in the discontinuity thus takes the following form, 
$\forall t \in \left[0,T\right]$:

\begin{equation}
\overrightarrow{OP}(t) = \left[
  \begin{array}{c}
    \sum_{i=0}^{5} a_i t^i \\
    \sum_{i=0}^{5} b_i t^i \\
    \sum_{i=0}^{5} c_i t^i 
  \end{array}
\right]
\label{eq:postion}
\end{equation}
\begin{equation}
\frac{d\overrightarrow{OP}(t)}{dt} = \left[
  \begin{array}{c}
    \sum_{i=1}^{5} i a_i t^{i-1} \\
    \sum_{i=1}^{5} i b_i t^{i-1} \\
    \sum_{i=1}^{5} i c_i t^{i-1}
  \end{array}
\right]
\label{eq:vitesse}
\end{equation}

\begin{equation}
\frac{d^2\overrightarrow{OP}(t)}{dt^2} = \left[
  \begin{array}{c}
    \sum_{i=2}^{5} \frac{i!}{\left( i-2\right)!} a_i t^{i-2} \\
    \sum_{i=2}^{5} \frac{i!}{\left( i-2\right)!} b_i t^{i-2} \\
    \sum_{i=2}^{5} \frac{i!}{\left( i-2\right)!} c_i t^{i-2}
  \end{array}
\right]
\label{eq:acceleration}
\end{equation}

\begin{equation}
\frac{d^3\overrightarrow{OP}(t)}{dt^2} = \left[
  \begin{array}{c}
    \sum_{i=3}^{5} \frac{i!}{\left( i-3\right)!} a_i t^{i-3} \\
    \sum_{i=3}^{5} \frac{i!}{\left( i-3\right)!} b_i t^{i-3} \\
    \sum_{i=3}^{5} \frac{i!}{\left( i-3\right)!} c_i t^{i-3}
  \end{array}
\right]
\label{eq:jerk}
\end{equation}

\subsection{Boundary conditions}

In order to resolve this system, eight boundary conditions are used:

\begin{eqnarray}
  \overrightarrow{OP}(0)& = & \overrightarrow{OM} \label{eq:cond1}\\
  \overrightarrow{OP}(T)& = &\overrightarrow{ON} \label{eq:cond2}\\
  \overrightarrow{OP}\left(\frac{T}{2}\right)& = &\overrightarrow{OQ} \label{eq:cond3}\\
  \overrightarrow{V}\left(0\right)& = &\overrightarrow{V_M} = -V_{In}\overrightarrow{u} \label{eq:cond4}\\ 
  \overrightarrow{V}\left(T\right)& = &\overrightarrow{V_N} = V_{In}\overrightarrow{v}\label{eq:cond5}\\
  \overrightarrow{A}\left(0\right)& = &\overrightarrow{0} \label{eq:cond6}\\
  \overrightarrow{A}\left(T\right)& = &\overrightarrow{0} \label{eq:cond7}\\
  \overrightarrow{J}\left(\frac{T}{2}\right)& = &\overrightarrow{0} \label{eq:cond8}
\end{eqnarray}

Equations \ref{eq:cond1} and \ref{eq:cond2} translate entry into the discontinuity. 
Equation \ref{eq:cond3} translates the problem's symmetry, meaning that the point parametrized by tolerances of passage is reached half way through the time of the path. Considering that acceleration is null at entry and exit of the discontinuity, equations \ref{eq:cond6} and \ref{eq:cond7} are obtained. To conclude, equation \ref{eq:cond8} translates symmetry of the acceleration contour.

\subsection{Resolution}
This thus involves resolving a system of 24 scalar equations (projection of equations \ref{eq:cond1} onto \ref{eq:cond8}) whose unknowns are:
\begin{itemize}
\item 18 coefficients $a_i$, $b_i$, $c_i$ of polynomials,
\item the norm L of vectors $\overrightarrow{OM}$ and
  $\overrightarrow{ON}$,
\item time $T$ for passage of the transition. 
\end{itemize}

Resolving the system gives: 

\begin{equation}
\left\{
  \begin{array}{l}
    a_0 = \frac{16 Q_x \sin\varphi_M\cos\theta_M}{3\left(
        \sin\varphi_N \cos\theta_N +  \sin\varphi_M  \cos\theta_M 
      \right)} \\
    b_0 = \frac{16 Q_y \sin\varphi_M\sin\theta_M}{3\left(
        \sin\varphi_N \sin\theta_N +  \sin\varphi_M  \sin\theta_M 
      \right)} \\
    c_0 = \frac{16 Q_z \cos\varphi_M }{3\left( \cos\varphi_N +
        \cos\varphi_M\right)} 
  \end{array}
\right.\label{eq:a0}
\end{equation}

\begin{equation}
\left\{
  \begin{array}{l}
    a_1 =-V_{In}\sin\varphi_M\cos\theta_M \\
    b_1 =-V_{In}\sin\varphi_M\sin\theta_M \\
    c_1 =-V_{In}\cos\varphi_M 
  \end{array}
\right.\label{eq:a1}
\end{equation}

\begin{equation}
\left\{
  \begin{array}{l}
    a_2 = 0 \\
    b_2 = 0 \\
    c_2 = 0
  \end{array}
\right.\label{eq:a2}
\end{equation}

\begin{equation}
\left\{
  \begin{array}{l}
    a_3 = \frac{9 \left(\sin\varphi_N \cos\theta_N+\sin\varphi_M
        \cos\theta_M\right)^3   V_{In}^3}{1024 Q_x^2} \\ 
    b_3 = \frac{9 \left(\sin\varphi_N \sin\theta_N+\sin\varphi_M
        \sin\theta_M\right)^3   V_{In}^3}{1024 Q_y^2} \\
    c_3 = \frac{9 \left(\cos\varphi_N+\cos\varphi_M \right)^3
      V_{In}^3}{1024 Q_z^2} 
  \end{array}
\right.\label{eq:a3}
\end{equation}

\begin{equation}
\left\{
  \begin{array}{l}
    a_4 = -\frac{27 \left(\sin\varphi_N \cos\theta_N+\sin\varphi_M
        \cos\theta_M\right)^4 V_{In}^4}{65536 Q_x^3} \\
    b_4 = -\frac{27 \left(\sin\varphi_N \sin\theta_N+\sin\varphi_M
        \sin\theta_M\right)^4 V_{In}^4}{65536 Q_y^3} \\
    c_4 = -\frac{27 \left(\cos\varphi_N+\cos\varphi_M \right)^4
      V_{In}^4}{65536 Q_z^3} 
  \end{array}
\right.\label{eq:a4}
\end{equation}

\begin{equation}
\left\{
  \begin{array}{l}
    a_5 = 0 \\
    b_5 = 0 \\
    c_5 = 0
  \end{array}
\right.\label{eq:a5}
\end{equation}

\begin{equation}
\begin{array}{c}
\overrightarrow{OP}(T) = \overrightarrow{ON}\Longleftrightarrow 
\\
\left(
\begin{array}{c}
T\\
T\\
T\\
\end{array}
\right)
=
\left(
\begin{array}{c}
   \frac{32 Q_x}{3 V_{In} \left( \sin\varphi_N \cos\theta_N +  \sin\varphi_M
       \cos\theta_M   \right)}   \\ 
   \frac{32 Q_y}{3 V_{In} \left( \sin\varphi_N \sin\theta_N +  \sin\varphi_M
       \sin\theta_M   \right)}   \\ 
   \frac{32 Q_z}{3 V_{In} \left( \cos\varphi_N  + \cos\varphi_M \right)}   \\
\end{array}
\right)
\end{array}
\label{eq:T}
\end{equation}

\begin{equation}
\begin{array}{c}
\begin{array}{c}
\overrightarrow{OP}\left(\frac{T}{2}\right) = \overrightarrow{OQ} 
\end{array}
\Longleftrightarrow
\\
\left(
\begin{array}{c}
L \\
L \\
L \\
\end{array}
\right)
=
\left(
  \begin{array}{c}
    \frac{16 Q_x}{3\left( \sin\varphi_N \cos\theta_N +  \sin\varphi_M
        \cos\theta_M \right)}  \\  
    \frac{16 Q_y}{3\left( \sin\varphi_N \sin\theta_N +  \sin\varphi_M
        \sin\theta_M  \right)}  \\
   \frac{16 Q_z}{3 \left( \cos\varphi_N  + \cos\varphi_M \right)}   \\ 
\end{array}
\right)
\end{array}
\label{eq:L}
\end{equation}

Due to the relation between $Q_x$, $Q_y$ and $Q_z$, the 3 expressions allowing $L$ or $T$ to be computed give the same results. With respect to the feed rate on entry of the discontinuity $V_{In}$, it is \textit{ex ante} equal to the feed rate at the end of the upstream block. Nevertheless, it can perhaps be limited by maximum jerk, maximum acceleration and the length of the blocks upstream and downstream from the transition. This feed rate can now be calculated.

\subsection{Feed rate on entry in the discontinuity}
\label{subsec:vindisc}
Expressing the problem as an equation allows the entry feed rate to be calculated when it is limited by acceleration or maximum jerk: the assumption is made that maximum jerk is at point $M$ and that acceleration will be at its maximum at $Q$. 
Care must therefore be taken to ensure that the machine's capabilities are not exceeded at these points. 

If the maximum jerk is reached on each of the axes, the equation \label{eq:jerk} gives:
\begin{equation}
\overrightarrow{J}(0) = \overrightarrow{J_{max,i}} \Longleftrightarrow
6
\left(
\begin{array}{c}
a_3\\
b_3\\
c_3
\end{array}
\right)
=
\left(
\begin{array}{c}
J _{max,x}\\
J _{max,y}\\
J _{max,z}
\end{array}
\right)
\end{equation}

According to the case, resolution of this system allows the entry rate limited by an axial jerk to be determined using the results given by equation \ref{eq:a3}:

\begin{equation}
  \begin{array}{lcl}
  V_{lim,j} &=& \frac{8}{3\left| \sin\varphi_N \cos\theta_N +
      \sin\varphi_M \cos\theta_M \right|} \cdot\min
  \Big(\\
&&    \sqrt[3]{Q_x^2 J_{max,x}},
    \sqrt[3]{Q_y^2 J_{max,y}},
    \sqrt[3]{Q_z^2 J_{max,z}}
  \Big)
  \end{array}
\label{eq:vlimj}    
\end{equation}

Similarly, it is known that on the discontinuity, acceleration is at its maximum on $T/2$. The entry feed rates that will be limited by acceleration can then be determined by resolving the following equation: 

\begin{equation}
\overrightarrow{A}\left(\frac{T}{2}\right) = \overrightarrow{A_{max,i}}
\end{equation}

By inserting this condition into equation \ref{eq:acceleration} and using the results given by equations \ref{eq:a3} and \ref{eq:a4}, the following is obtained:

\begin{equation}
\begin{array}{lcl}
  V_{lim,a}& = &\frac{8}{3\left| \sin\varphi_N \cos\theta_N +
      \sin\varphi_M \cos\theta_M \right|} \min
  \Big(\\
    &&\sqrt{Q_x A_{max,x}},
    \sqrt{Q_y A_{max,y}},
    \sqrt{Q_z A_{max,z}}
  \Big)
\end{array}
\label{eq:vlima}
\end{equation}

Thus, maximum acceleration on each of the axes also leads to limitations on the feed rate on entering the block: 
\begin{equation}
  \label{eq:vin}
  V_{In} = \min\left(V_F, V_{lim,j},V_{lim,a} \right)
\end{equation}

Finally, a last case remains: the transition length calculated $L$ can be greater than the length of the segment upstream or downstream. An additional condition is thus imposed: if $L>||\overrightarrow{OA}||/2$  or $L>||\overrightarrow{OB}||/2$, the coordinates of point $Q$ are recalculated taking $L=\min\left(||\overrightarrow{OA}||/2,||\overrightarrow{OB}||/2
\right)$ from equation \ref{eq:L}. Equations
\ref{eq:vlimj} and \ref{eq:vlima} then allow the limit feed rate on entry in the discontinuity to be recalculated.

The proposed model is now complete. Experimental validation is proposed in the following section. 

\section{Experimental validation}
\label{sec:validation}

Measurements of position profiles, feed rates and acceleration were
made on a machine with the aim of validating modelling.  
Tests were conducted on a DMU 50 eVo 5-axis machining centre equipped
with a Siemens 840D Numerical Control. The characteristics of the NC
and machine combination are given in table
\ref{tab:caracteristiques}. This NC allows measurements of position,
feed rate and acceleration to be made for a maximum period of 10
seconds.  

\begin{table}[!ht]
  \centering
  \caption{Characteristics DMU 50 eVo - Siemens 840 D}
  \begin{tabular}{cc}
    \hline
    \multicolumn{2}{c}{Dynamic characteristics of axes}\\
    \hline
    Maximum feed rate & $V_{max,i} = 50 \; m/min$  \\
    Maximum acceleration & $A_{max,i} =9,8\; m/s^2$ \\
    Maximum jerk in translation & $J_{max,i} = 40\; m/s^3$\\
    Maximum jerk on passage of & 
    $Jc_{max} =60\;  m/s^3$  \\ 
    a discontinuity in curvature  &\\
    \hline
    \multicolumn{2}{c}{Characteristics of the NC}\\
    \hline
    Tolerances on the axes  & $Tol = 0,01\; mm$ \\
    Interpolation time of NC & $2 ms$ \\
    \hline
  \end{tabular}
  \label{tab:caracteristiques}
\end{table}

\begin{figure*}[!ht]
  \caption{Simulated paths \label{fig:spiraleg1g2}}
  \begin{tabular}{cccc}
    \begin{tabular}{c}
      \includegraphics[width=0.25\textwidth]{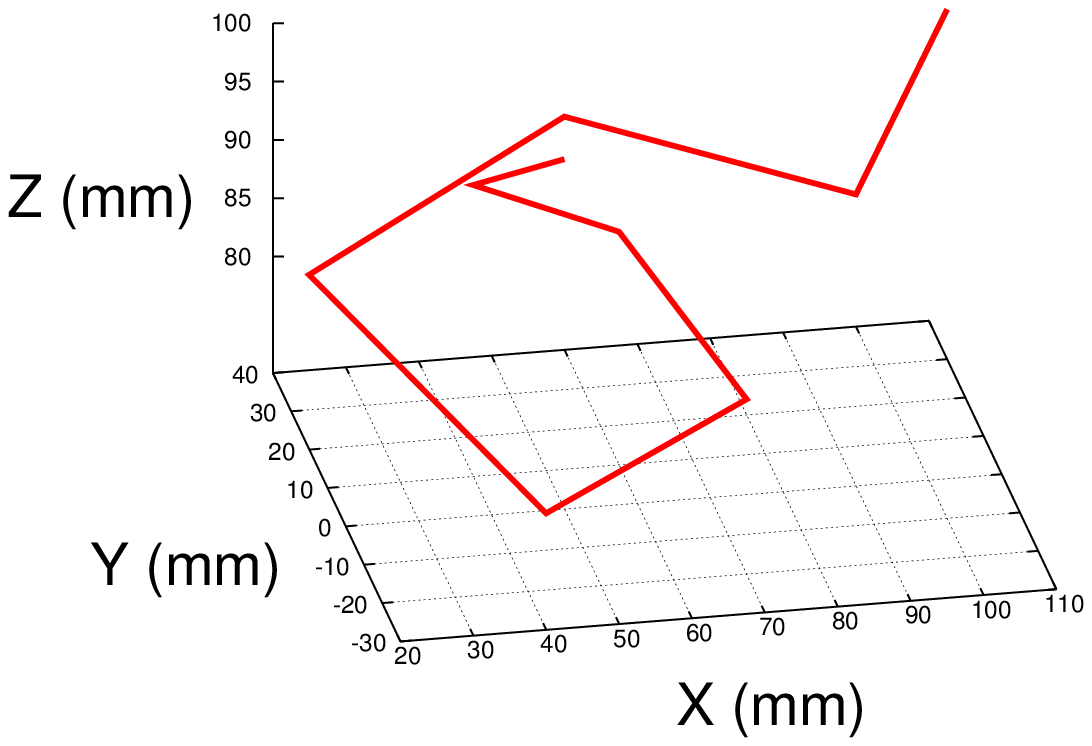}
    \end{tabular}
    &
    \footnotesize{
      \begin{tabular}{ccc}
        \hline
        X   & Y   & Z   \\
        \hline
        50  & 0   & 100 \\
        40  & 10  & 95  \\
        60  & 10  & 90  \\
        70  & -20 & 85  \\
        40  & -30 & 80  \\
        20  & 20  & 85  \\
        60  & 40  & 90  \\
        90  & 0   & 95  \\
        110 & 30  & 100 \\
        \hline
      \end{tabular}}
    &
    \begin{tabular}{c}
      \includegraphics[width=0.25\textwidth]{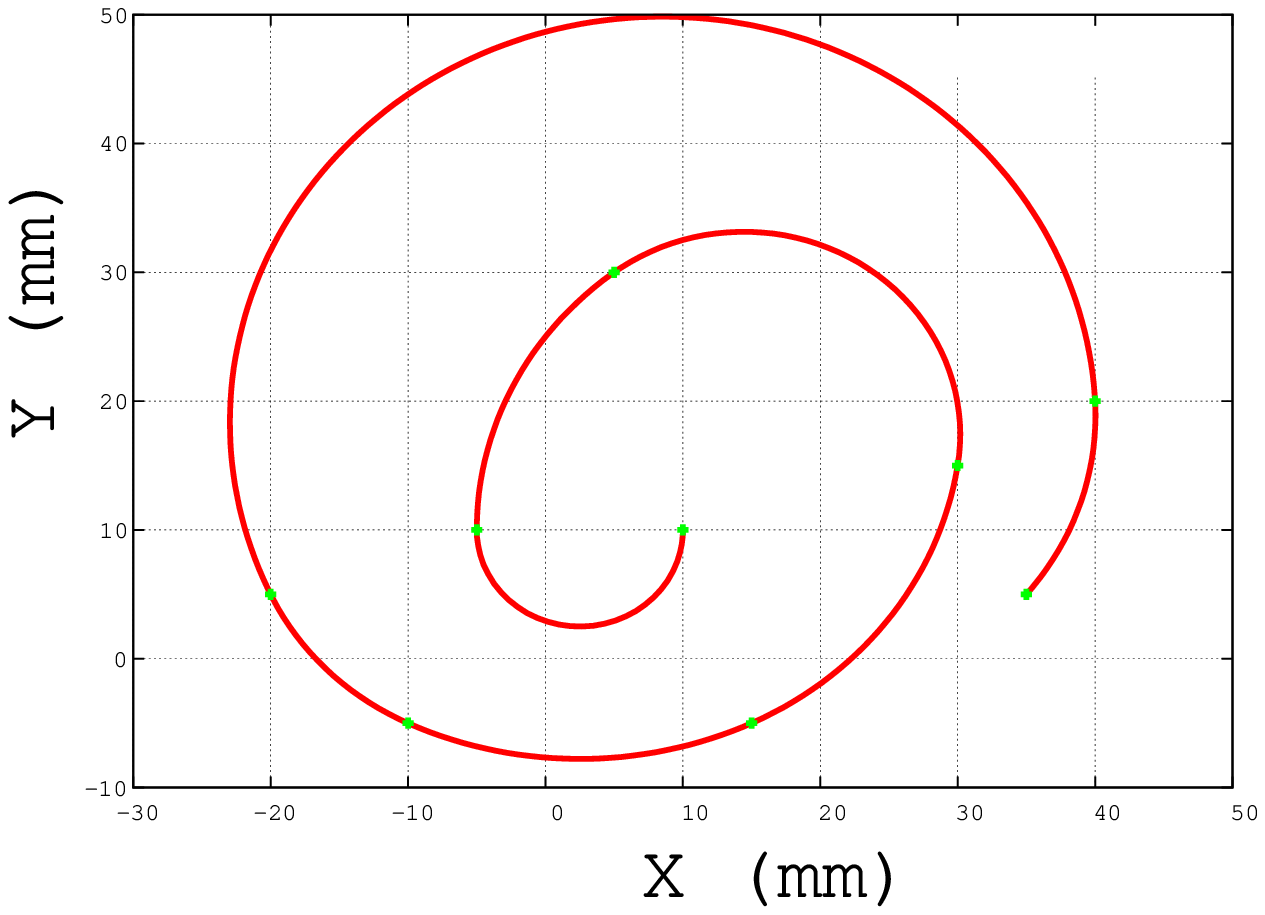}
    \end{tabular}
    &\footnotesize{
      \begin{tabular}{ccc}
        \hline
        X   & Y     & R \\
        \hline
        -5.0  & 10.0 & 7.5   \\
        5.0   & 30.0 & 25    \\
        30.0  & 15.0 & 15.74 \\
        15.0  & -5.0 & 26.56 \\
        -10.0 & -5.0 & 29.51 \\
        -20.0 & 5.0  & 20.73 \\
        40.0  & 20.0 & 31.50 \\
        35.0  & 5.0  & 21.62 \\
        \hline
      \end{tabular}
    } \\
    &&&\\
    \multicolumn{2}{c}{3-axis paths in linear  interpolation}& 
    \multicolumn{2}{c}{2-axis paths in circular interpolation} 
  \end{tabular}
\end{figure*}

A test campaign enabled the proposed models to be validated. In
particular, feed rates ranging from 500 to 10000 $mm/min$ and
machining tolerances from 0.1 to 0.01 $mm$ on various paths were
tested. Moreover, as the NC allowed certain dynamic parameters to be
modified, jerk was also subjected to variation. In the present
publication, only two tests are presented (figure
\ref{fig:spiraleg1g2}\footnote{In the case of the circular
  interpolation, X and Y are the coordinates of the
          final point}): one test in linear interpolation and a second one in circular interpolation. The first test sought to validate the simulation method adopted as well as the method for passage of discontinuities using polynomials. The second confirmed the modelling of behavioural laws on circular portions. The programmed feed rate was $5000\;mm/min$ on the cases studied.  

\begin{figure}[!ht]
  \centering
  \caption{Measured and simulated contours for cutter - workpiece feed rates \label{fig:vitessesg1g2}}    
  \begin{tabular}{c}
    \includegraphics[width=.4\textwidth]{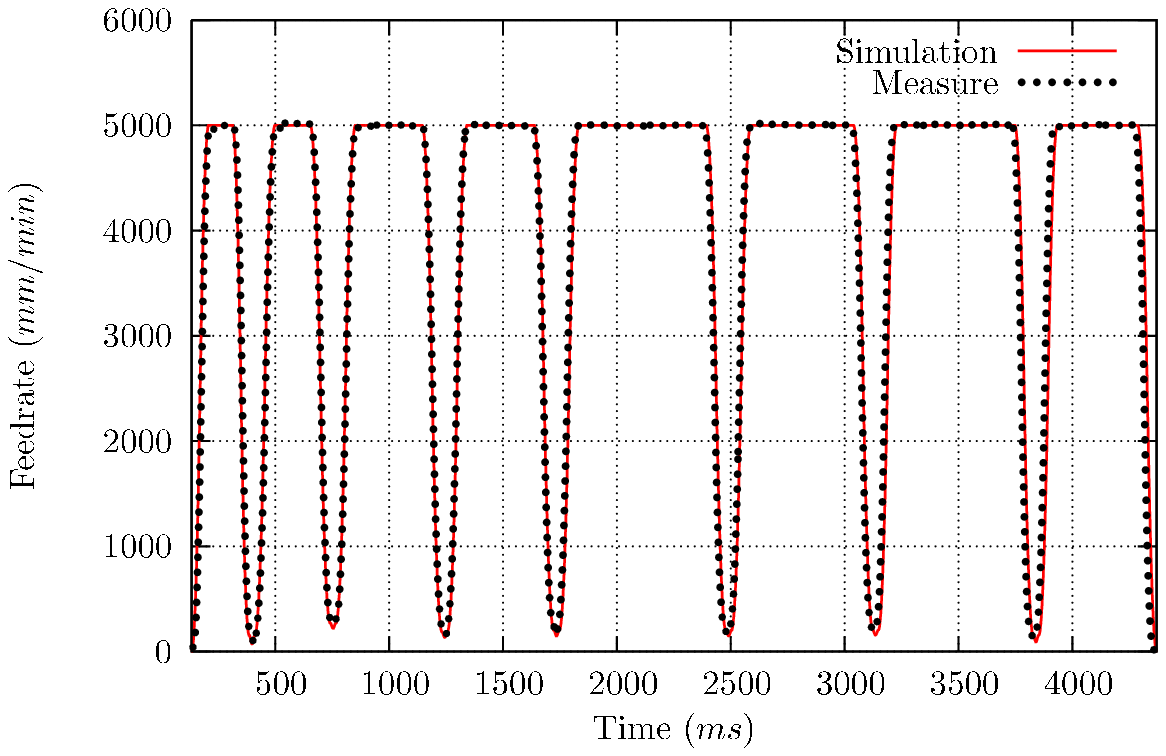} \\
    Linear  interpolation \\
    \includegraphics[width=.4\textwidth]{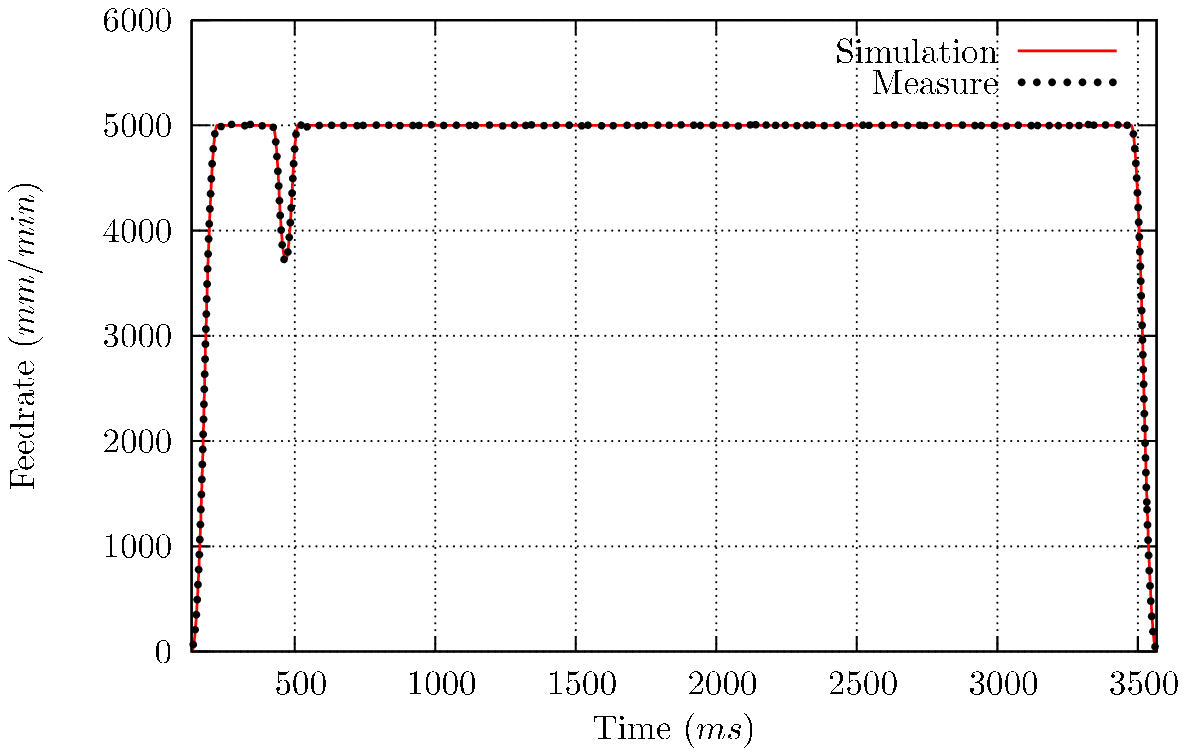} \\
    Circular interpolation \\
  \end{tabular}
\end{figure}

The cutter - workpiece feed rate profiles measured and simulated were plotted on figure \ref{fig:vitessesg1g2}. What is immediately striking is the model's faithfulness to the experimental curves.

In linear interpolation, the path time calculated is $4 237\;
ms$ and the time measured on the machine is $4232\; ms$
\textit{i.e.} an error of less than 0.2\%. Error between simulation and
measurement reaches a maximum level on drops in feed rate at passages
between discontinuities (table \ref{tab:erreurg1}). Although errors
between the rate measured and the rate simulated in the discontinuity
can appear to be significant, such error on the drop can be seen to
represent less than 1\%, the profiles are seen to be closed and error on the time of pass through a discontinuity is low.

\begin{table*}[!ht]
  \centering
  \caption{Errors calculated on the cutter - workpiece feed rate on passages of discontinuities (Feed rate in $mm/min$)\label{tab:erreurg1}}
  \begin{tabular}{cccccccc}
    \hline
    & \multicolumn{3}{c}{Feed rate in the transition} 
    && \multicolumn{3}{c}{Drop in feed rate }\\
    & Measurement & Simulation & Error && Measurement & Simulation & Error \\
    \hline
    1 & 71  & 72  & 1,4  \% && 4929 & 4928 & 0,02 \%\\
    2 & 256 & 222 & 13,28 \% && 4744 & 4778 & 0,7  \%\\
    3 & 150 & 135 & 10,67 \% && 4850 & 4865 & 0,31 \%\\
    4 & 193 & 148 & 23,31 \% && 4807 & 4852 & 0,94 \%\\
    5 & 161 & 150 & 6,83  \% && 4839 & 4850 & 0,23 \%\\
    6 & 196 & 158 & 19,39 \% && 4804 & 4842 & 0,79 \%\\
    7 & 126 & 91  & 27,78 \% && 4874 & 4909 & 0,72 \%\\
    \hline
  \end{tabular}
\end{table*}

\begin{figure}[!ht]
  \centering
  \caption{Comparison of contours measured and simulated on a rectilinear path \label{fig:xpva}}    
  \begin{tabular}{c}
    \includegraphics[width=.4\textwidth]{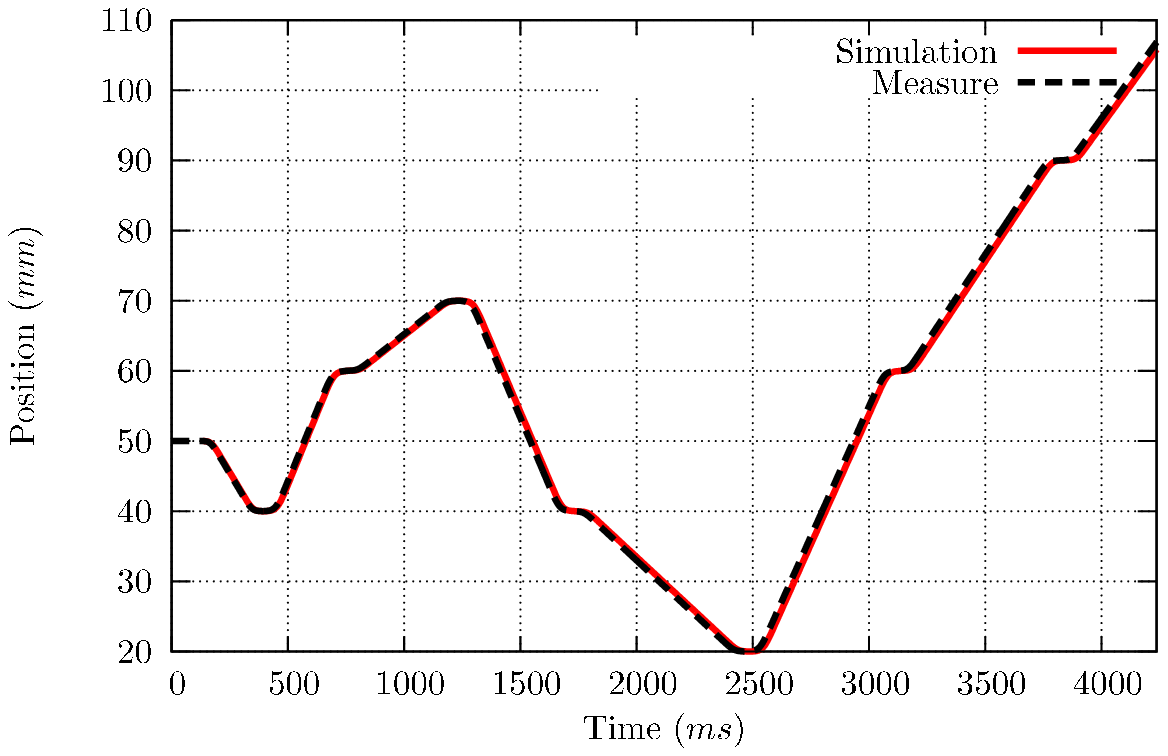} \\
    Position \\
    \includegraphics[width=.4\textwidth]{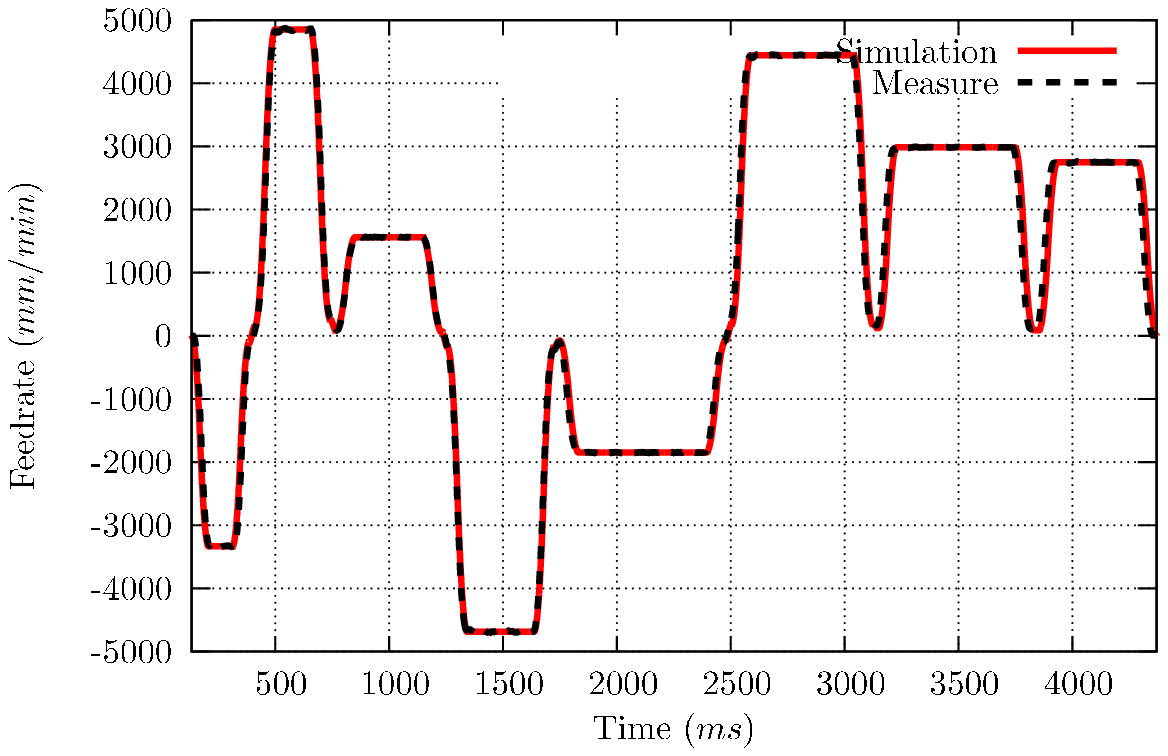} \\
    Feed rate \\
    \includegraphics[width=.4\textwidth]{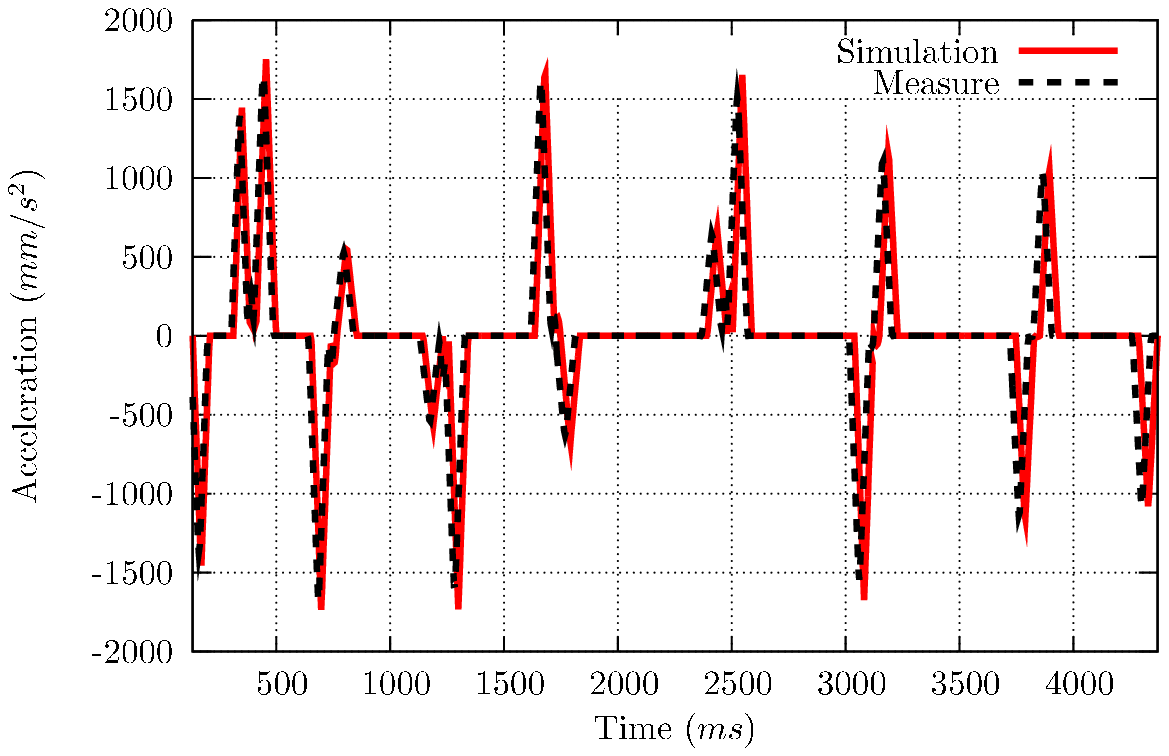} \\
    Acceleration 
  \end{tabular}
\end{figure}

\begin{figure}[!ht]
  \centering
  \caption{Passage of a discontinuity at a tangent \label{fig:xpvaloc}}    
  \begin{tabular}{c}
    \includegraphics[width=.4\textwidth]{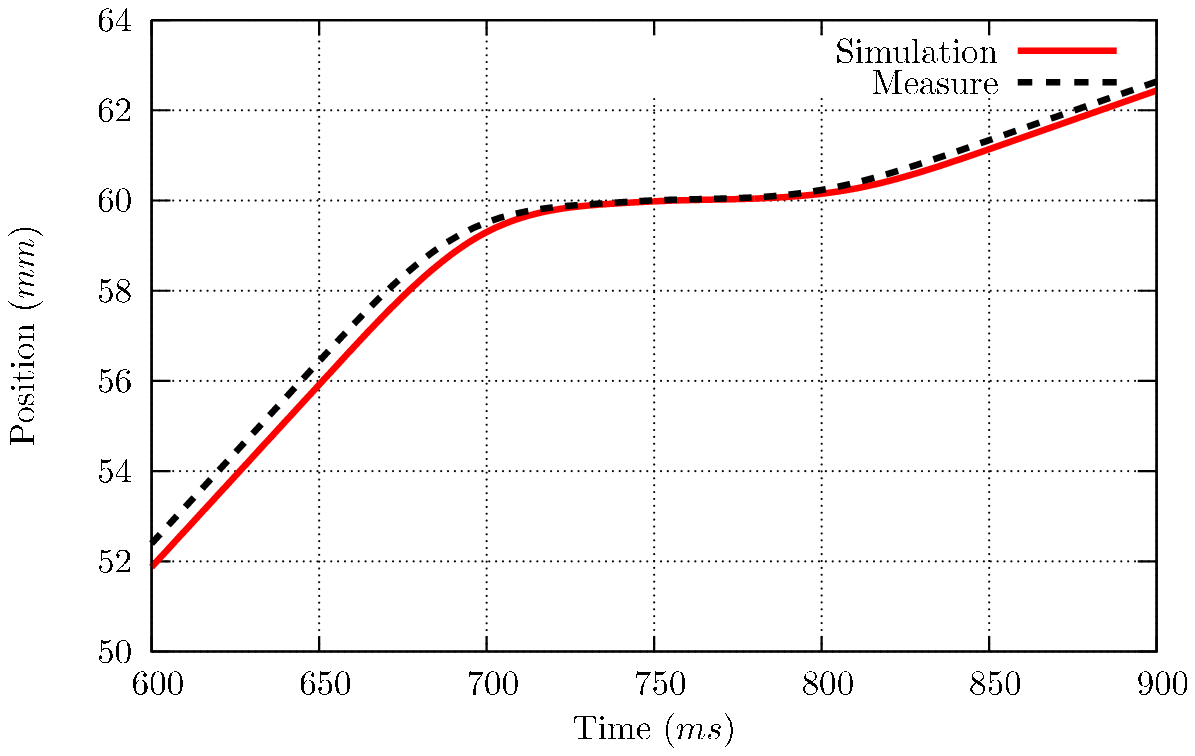} \\
    Position \\
    \includegraphics[width=.4\textwidth]{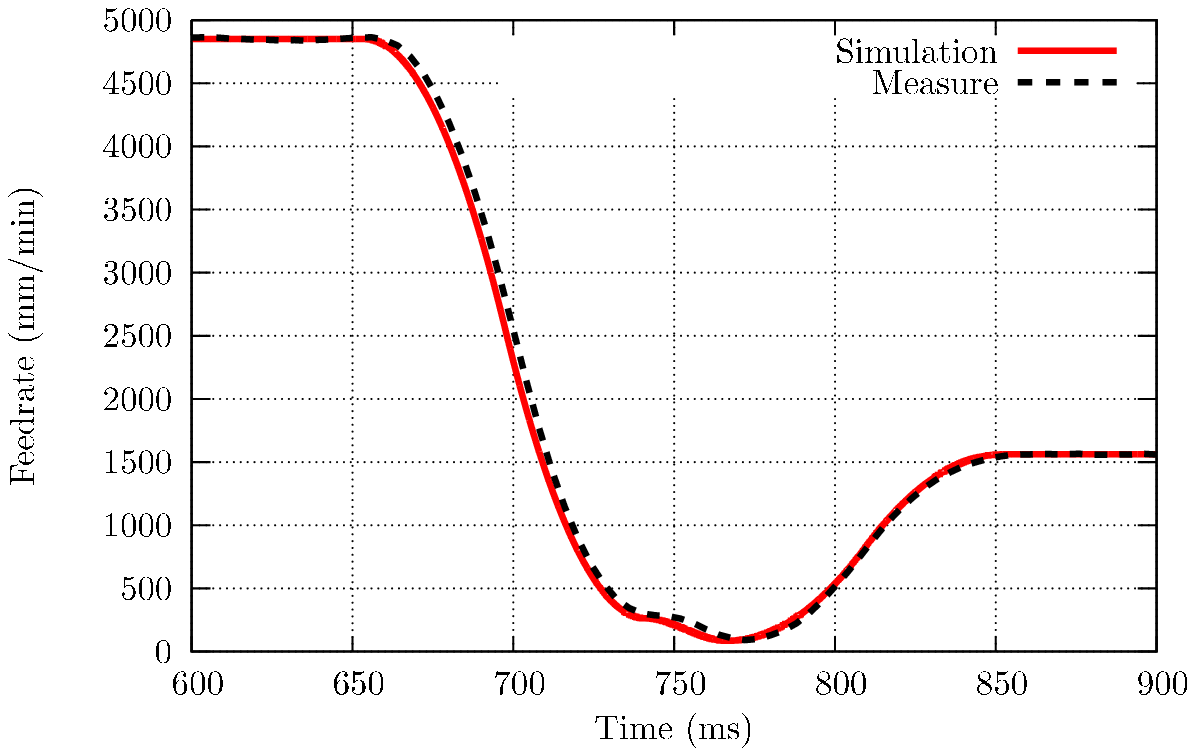} \\
    Feed rate \\
    \includegraphics[width=.4\textwidth]{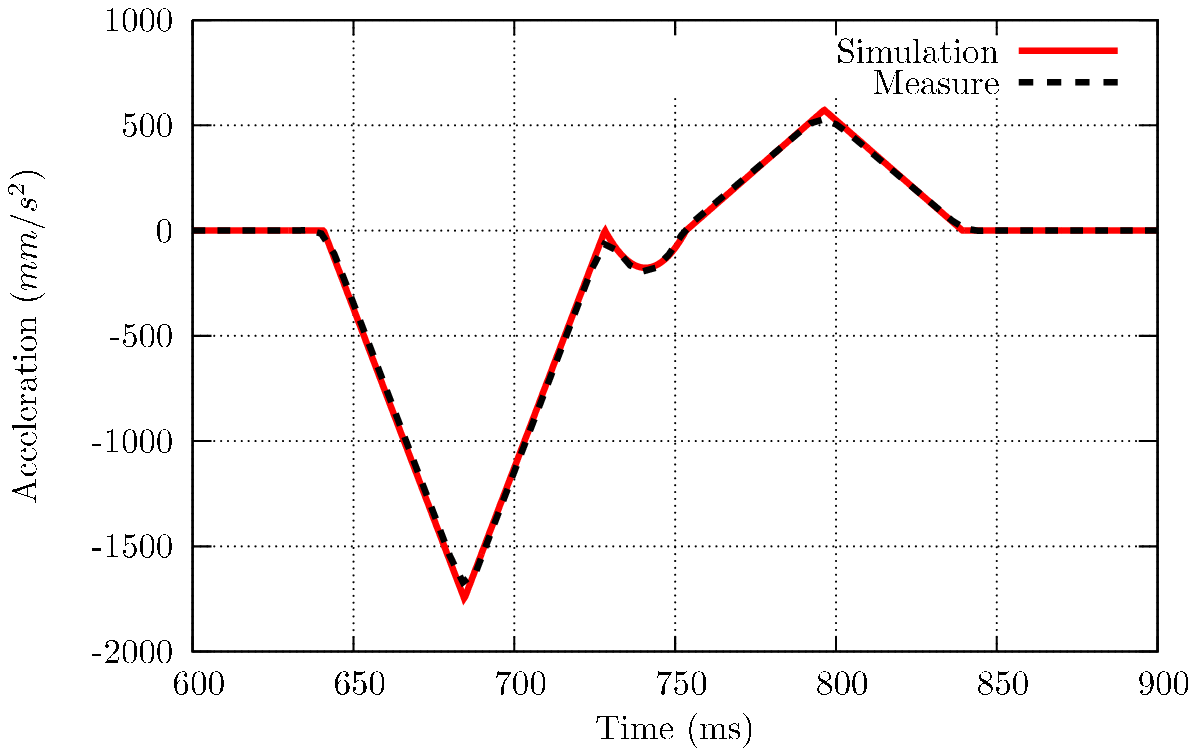} \\
    Acceleration 
  \end{tabular}
\end{figure}

Figure \ref{fig:xpva} shows the plots of positions, feed rates and accelerations in projection on the $\overrightarrow{X}$ axis. An excellent match can be seen between the model and the measurement. Figure
\ref{fig:xpvaloc} represents a zoom onto the zone on crossing the discontinuity between the 600th and 900th milliseconds to pass
points 2 -- 3 -- 4. Note that on the measurements, the feed rate diminishes on the segment to reach a rate of $281\;mm/min$. The calculated feed rate is $263\; mm/min$, representing an error of 6.8\%. The discontinuity is then entered. The feed rate measured at the end of the discontinuity is $91\; mm/min$. The rate calculated is $85\; mm/min$ i.e. 7\% of error. Finally, observation of measurement of the acceleration curve shows that there is no leap in acceleration on entering the discontinuity as would tend to suggest passage of a transition by a circle arc. This corroborates the validity of using the model in which acceleration remains continuous.

In circular interpolation, the radius values chosen only highlight a drop in feed rate on passage of the first discontinuity. Error between the calculated value and the simulated value is here only 2\%. Error on the difference between the total simulated time and the total calculated time is here also less than 1\%.

\begin{figure}[!ht]
  \centering
  \caption{Comparison of contours measured and simulated on the circular path \label{fig:g2}} 
  \begin{tabular}{c}
    \includegraphics[width=.4\textwidth]{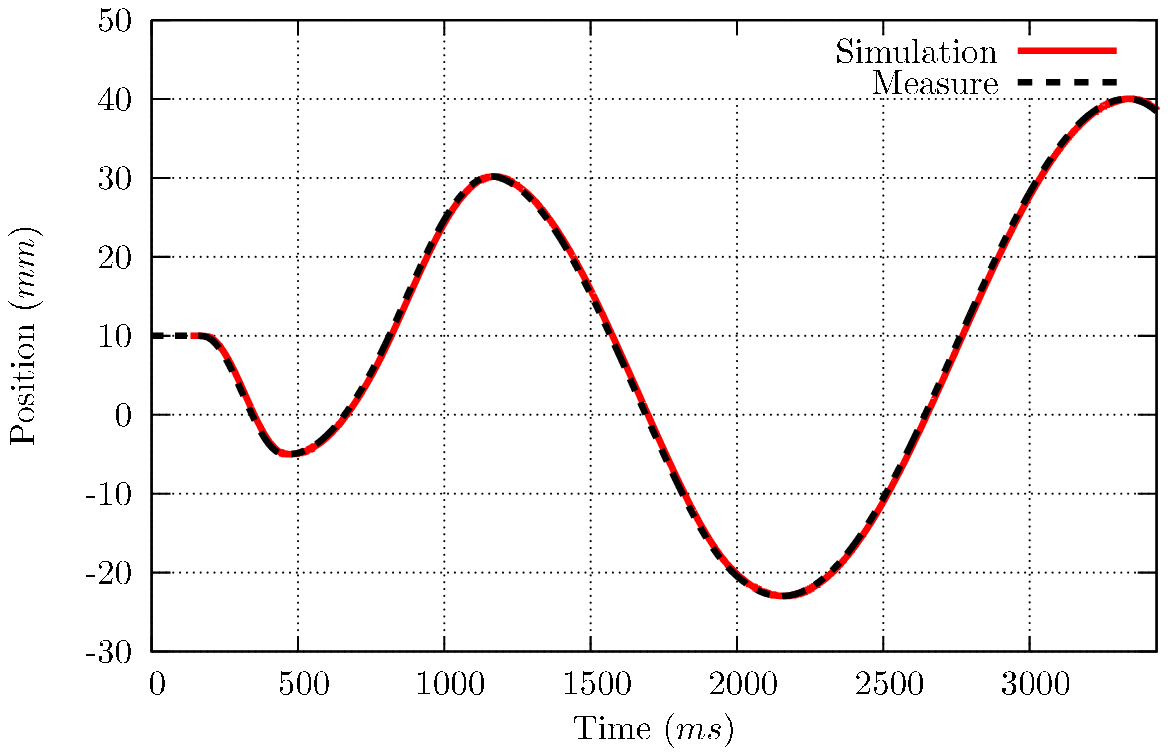} \\
    Position \\
    \includegraphics[width=.4\textwidth]{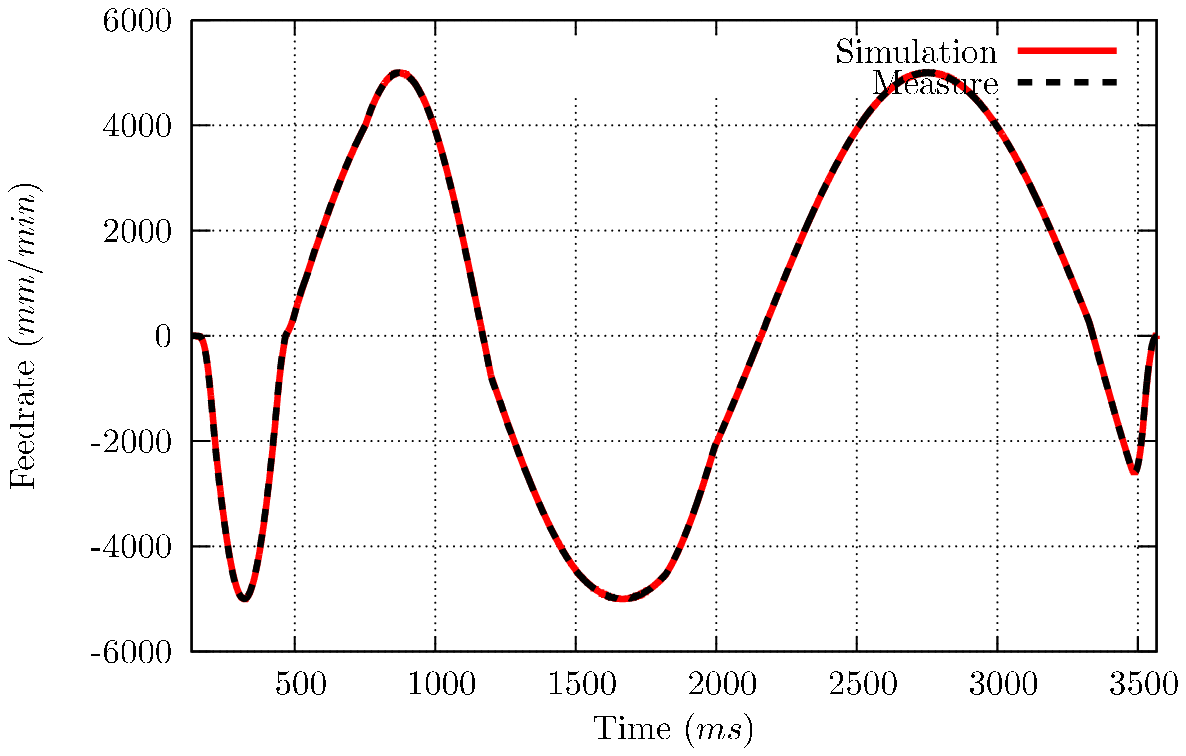} \\
    Feed rate \\
    \includegraphics[width=.4\textwidth]{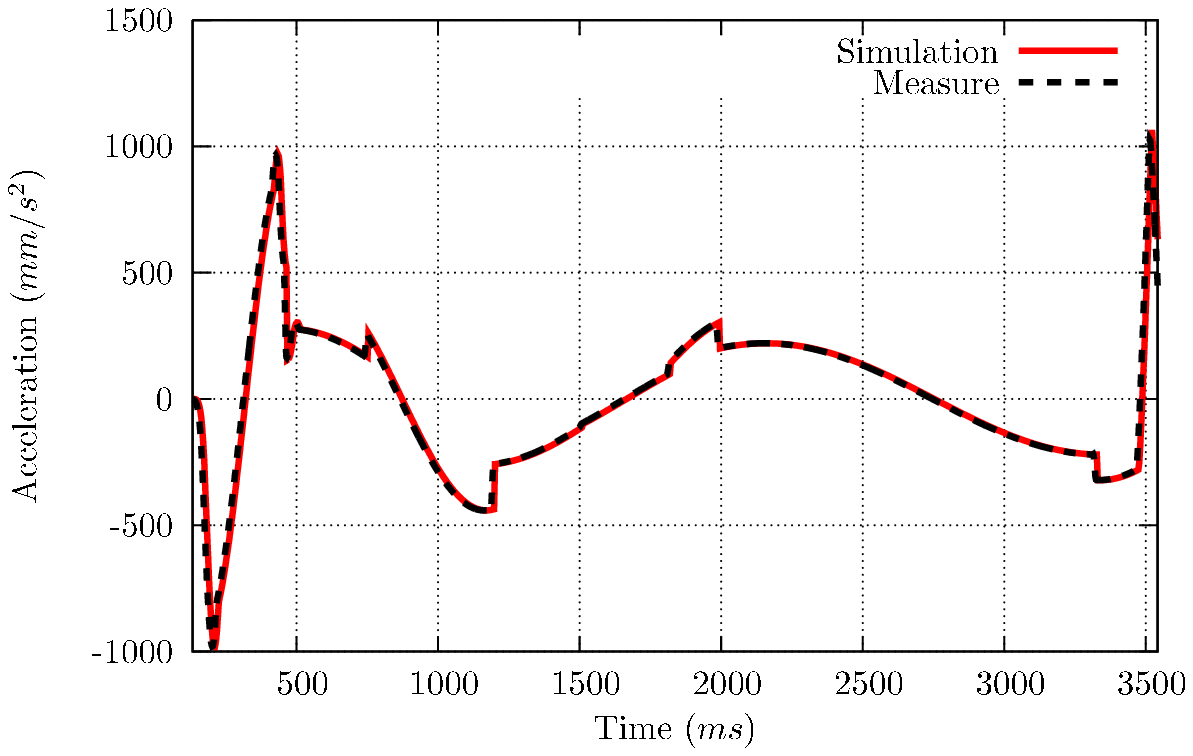} \\
    Acceleration 
  \end{tabular}
\end{figure}

Figure \ref{fig:g2} shows positions, feed rates and accelerations
measured and simulated on the spiral programmed in circular
interpolation, in projection on the $\overrightarrow{X}$ axis. The
acceleration profile perfectly illustrates the leaps on passage of
discontinuities in curvature. The Pateloup model used for simulation
is thus fully confirmed. 

\begin{figure}[!ht]
  \centering
  \caption{Passage of a discontinuity in a curvature\label{fig:locg2}}
  \begin{tabular}{c}
    \includegraphics[width=.4\textwidth]{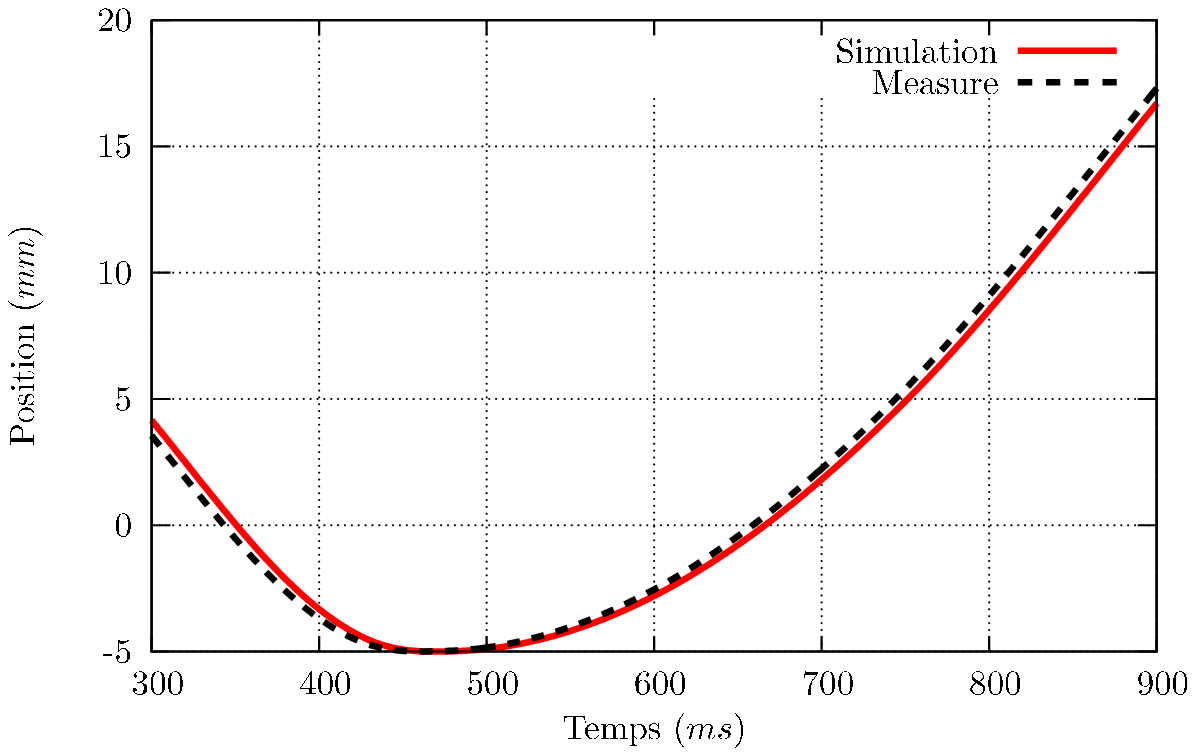} \\
    Position \\
    \includegraphics[width=.4\textwidth]{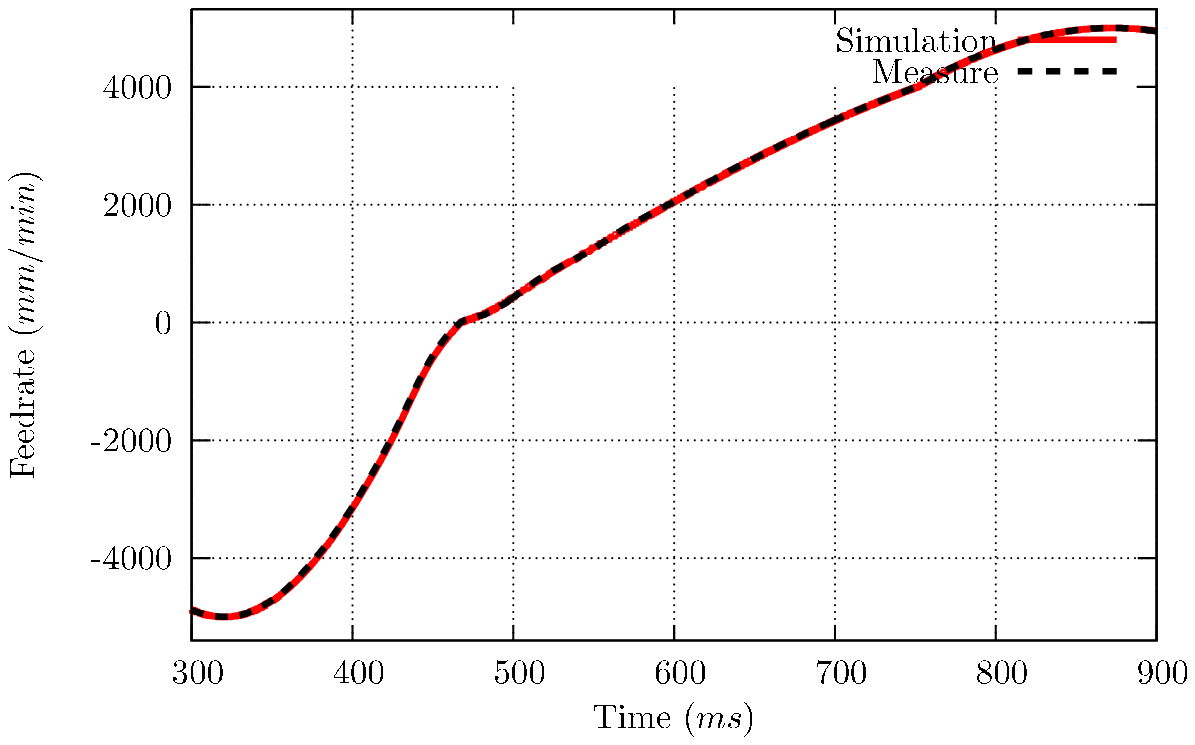} \\
    Feed rate \\
    \includegraphics[width=.4\textwidth]{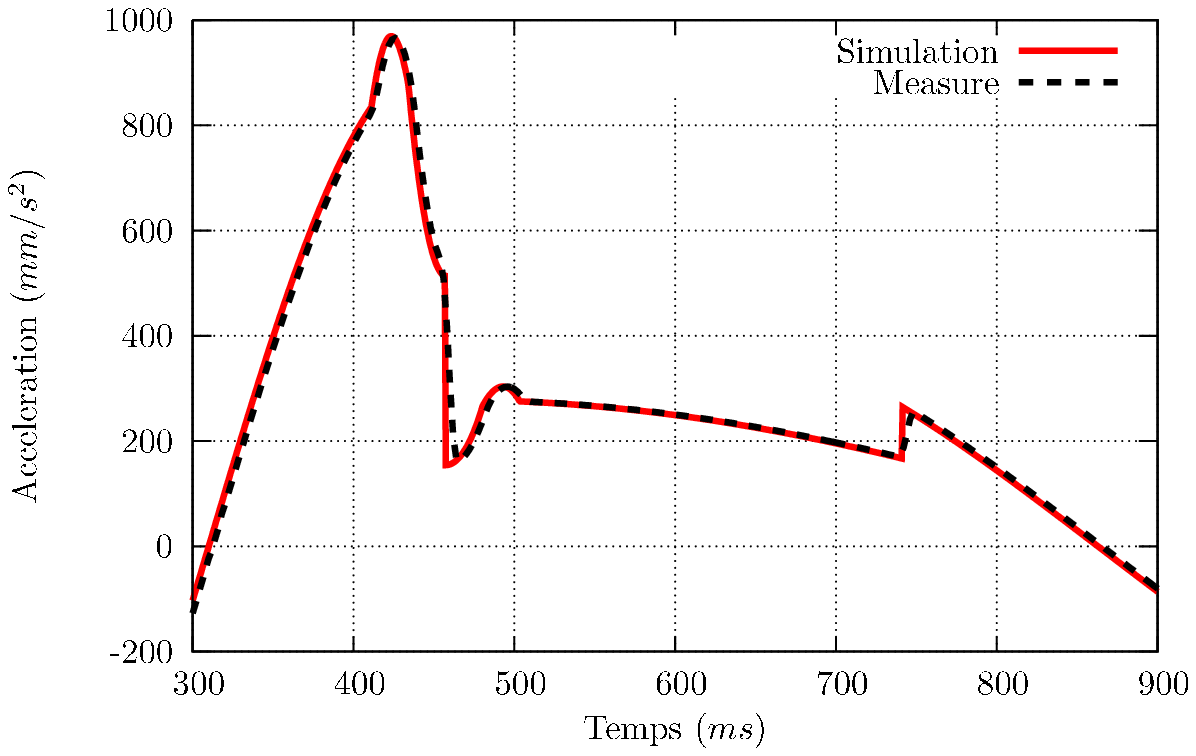} \\
    Acceleration
  \end{tabular}
\end{figure}

Figure \ref{fig:locg2} represents a zoom on a passage through discontinuity. The model can be seen to faithfully reflect the measurement. 

\section{Conclusion}
The present publication proposes a model allowing jumps of discontinuity in tangency to be overcome together 
with a model to simulate the axial behaviour of a machine in circular interpolation. This means that the position, 
feed rate, acceleration and jerk on each of the axes catering for the dynamic parameters of 
the machine/NC combinations can be simulated on 3 axes and cutter - workpiece contours be determined. This gives faithful simulation of the machine's dynamic
 behaviour in 3-axis machining in the commonest programming modes. 

Moreover, the present study stresses the problem of using linear interpolation when finishing complex workpieces.
 In this context, programmed tolerances are often extremely low and segments thus extremely short, leading to repeated reductions in feed rate. 
This problem is only very partially resolved by NCs using predictive functions. 

This means that beyond a certain segment size, the type of programming needs to be changed. Future works will therefore study 
the machine dynamics in polynomial interpolation and integration of this type of programming into the simulator. The possibility 
of integrating the behaviour of series 5-axis machines into the model can then be considered. 
Finally, in the longer term, the simulator should allow for machining paths to be optimized so as to boost the productivity of 
ultra-high speed machines.

\subsection*{Acknowledgements}

This work was carried out within the context of the working group Manufacturing 21 which brings together 11 French research laboratories. The topics addressed are as follows:
\begin{itemize}
\item modelling of the manufacturing process,
\item virtual machining,
\item emerging manufacturing methods. 
\end{itemize}

\section*{Appendices: General case for the path of a uniaxial segment}
\label{sec:annexes}

The non-linear system described in section \ref{sec:int} therefore needs to be resolved. This resolution involves resolving the 10 cases shown in figure \ref{fig:algocas}. 

The assumption is made that acceleration is null on entry and exit of the block. This is verified when the segments are long enough to reach the feed rate programmed at the end of the segment (section \ref{subsec:vindisc}).

Resolution of this system means the times for each of the phases can be determined so as to plot the cutter - workpiece feed rate profile. In some cases, the programmed feed rate will not be reached and the rate actually reached will be sought. 

According to the distance to be covered and the jumps in feed rate to be crossed, 10 cases were listed. From resolution of equations
\ref{eq:phase1} to \ref{eq:phase7}, the following obtains:

\begin{itemize}
\item Case 1: $V_{Out}$ cannot be reached. $V_{Out}'$ is reached by
  reaching $A_{max,i}$ (figure \ref{fig:case12}):
  \begin{equation}
    \left\{
      \begin{array}{l}
        |V_{Out} - V_{In} |\geq \frac{A_{max,i}^2}{J_{max,i}} \\
        2 V_{In} \frac{A_{max,i}}{J_{max,i}} +
        \frac{A_{max,i}^3}{J_{max,i}^2} \leq L \leq \\
        \quad \quad \frac{J_{max,i}
        (V_{Out}^2 - V_{In}^2) + A_{max,i}^2 (V_{In} +V_{Out})}{2
        A_{max,i} J_{max,i}}
    \end{array}
    \right.
    \label{eq:1.cond}
  \end{equation}
  \begin{equation}
    \begin{array}{lcl}
    V_{Out}'&=&\frac{1}{2 J_{max,i}} \\
    &&\cdot
    \Big( A_{max,i} J_{max,i}^2 L+4 J_{max,i}^2 V_{In}^2\\
    && -4 A_{max,i}^2 J_{max,i} V_{In}+A_{max,i}^4\Big)^{1/2}\\
    && -A_{max,i}^2
      
    \end{array}
    \label{eq:1.res1}
  \end{equation}
  \begin{equation}
    \left\{
      \begin{array}{l}
        \tau_1 = \tau_3 = \frac{A_{max,i}}{J_{max,i}} \\
        \tau_2 = \frac{V_{Out}'-V_{In}}{A_{max,i}} -
        \frac{A_{max,i}}{J_{max,i}}\\
        \tau_4 = \tau_5 = \tau_6 = \tau_7 = 0
      \end{array}
    \right.
    \label{eq:1.res2}
  \end{equation}
\item Case 2: $V_{Out}$ cannot be reached. $V_{Out}'$ is reached
  without reaching $A_{max,i}$ (figure \ref{fig:case12}):
  \begin{equation}
    \label{eq:2.cond1}    
    \left\{
      \begin{array}{l}
        |V_{Out} - V_{In}| \geq  \frac{A_{max,i}^2}{J_{max,i}} \\
        L\leq 2 V_{In} \frac{A_{max,i}}{J_{max,i}} +
        \frac{A_{max,i}^3}{J_{max,i}^2}
      \end{array}
    \right.
  \end{equation}
  \begin{equation}
    \left\{
      \begin{array}{l}
        |V_{Out} - V_{In}| \leq \frac{A_{max,i}^2}{J_{max,i}} \\
        L \leq \left(V_{Out} +V_{In} \right) \sqrt{\frac{V_{Out}
            -V_{In}}{J_{max}}} 
      \end{array}
    \right.
    \label{eq:2.cond2}
  \end{equation}
  
  \begin{equation}
    \label{eq:2.res1}
    V_{Out}' = V_{In} + J_{max,i}\tau_1^2
  \end{equation}
  \begin{equation}
    \label{eq:2.res2}
    \left\{
      \begin{array}{l}
        \tau_1 \textrm{ is solution of }    J_{max,i}
        \tau_1^3+2V_{In}\tau_1-L=0 \\
        \tau_1 = \tau_3 \\
        \tau_2 = \tau_4 = \tau_5 = \tau_6 = \tau_7 = 0
      \end{array}
    \right.
  \end{equation}

\begin{figure}[!ht]
  \caption{Cases 1 and 2 \label{fig:case12}}    
  \centering
  \begin{tabular}{c}
    \includegraphics[width=0.4\textwidth]{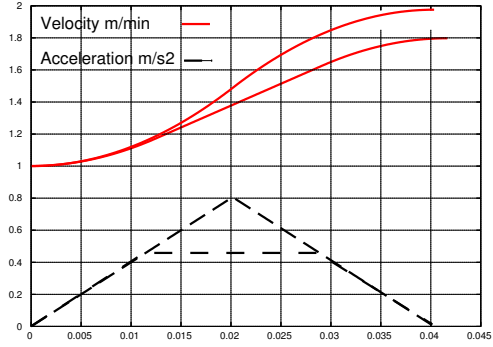}\\
  \end{tabular}
\end{figure}

\item Case 3: $V_{Out}$ is reached by reaching $A_{max,i}$. $V_F$
  is not reached. $V_F'$ is reached by reaching
  $A_{max,i}$. $\tau_1 = \tau_3= \tau_5 = \tau_7 =
  A_{max,i}/J_{max,i}$ $V_F'$, $\tau_1$ and 
  $\tau_6$ are solutions of: 
  \begin{equation}
    \label{eq:3.res}
    \left\{
      \begin{array}{lcl}
        V_F' &=& \frac{A_{max,i}^2}{J_{max,i}}+V_{In}+A_{max,i} \tau_2
        \\
        V_{Out} &=& -\frac{A_{max}^2}{J_{max,i}}+V_F'-A_{max,i} \tau_6 \\
        L &=&
        \frac{1}{2J_{max,i}} \cdot \\
        &&\Big(\big(2\tau_2V_{In}+2\tau_6V_F'-A_{max,i}\tau_6^2 \\
        &&+ A_{max,i}\tau_2^2\big) \cdot J_{max,i}\\
        && +4A_{max,i}V_{In} + 4A_{max,i}V_F'\\
        && -3A_{max,i}^2\tau_6+3A_{max,i}^2\tau_2 \Big)
      \end{array}
    \right.
  \end{equation}

\item Case 4: $V_{Out}$ is reached without reaching $A_{max,i}$. $V_F$
  is not reached. $V_F'$ is reached by reaching $A_{max,i}$. 
  $\tau_1 = \tau_3=  A_{max,i}/J_{max,i}$ and $V_F'$, $\tau_2$,
  $\tau_5$ are solutions of:
 \begin{equation}
    \label{eq:4.res}
    \left\{
      \begin{array}{lcl}
        V_F' &=& \frac{A_{max,i}^2}{J_{max,i}}+V_{In}+A_{max,i} \tau_2 \\
        V_{Out} &=& V_F'-\tau_5^2 J_{max,i}\\
        L &=& -\tau_5^3 J_{max,i}+\frac{2 A_{max} V_{In}}{J_{max,i}}
        \\
        && + \frac{3 A_{max,i}^2 \tau_2}{2 J_{max,i}}+
        \frac{A_{max,i}^3}{J_{max,i}^2} \\
        && +\tau_2 V_{In}+2 \tau_5
        V_F'+\frac{A_{max,i} \tau_2^2}{2} 
      \end{array}
    \right.
  \end{equation}

\item Case 5: $V_{Out}$ is reached by reaching $A_{max,i}$. $V_F$
  is not reached. $V_F'$ is reached without reaching $A_{max,i}$. 
  $\tau_5 = \tau_7 = A_{max,i}/J_{max,i}$, $\tau_2 = 0$,
  $\tau_1=\tau_3$, $\tau_1$, $\tau_6$ and $V_F'$ are solutions of:
    \begin{equation}
    \label{eq:5.res}
    \left\{
      \begin{array}{lcl}
        V_F' &=& V_{In} + J_{max,i} \tau_1^2 \\
        V_{Out} &=& -\frac{A_{max,i}^2}{J_{max,i}}+V_F'-A_{max,i} \tau_6
        \\
        L &=& \frac{1}{2 J_{max,i}^2}\cdot\\
        && \Big( 2 \tau_1^3 J_{max,i}^3\\
        &&+(4 \tau_1 V_{In}+ 2 \tau_6
        V_F'-A_{max,i} \tau_6^2) J_{max,i}^2 \\ 
        &&+(4 A_{max,i} Vc-3 A_{max,i}^2 \tau_6)
        J_{max,i}\\
        &&-2 A_{max,i}^3  \Big)
      \end{array}
    \right.
  \end{equation}

\item Case 6: $V_{Out}$ is reached without reaching $A_{max,i}$. $V_F$
  is not reached. $V_F'$ is reached without reaching $A_{max,i}$.
  $\tau_1 = \tau_3$, $\tau_5 = \tau_7$, $\tau_1$, $\tau_5$ and $V_F'$
  are solutions of (figure \ref{fig:case6}) :
    \begin{equation}
    \label{eq:6.res}
    \left\{
      \begin{array}{l}
        V_F' = \tau_1^3 J_{max,i}+2 \tau_1 V_{In} \\
        V_{Out} = V_F'-\tau_7^2 J_{max,i}  \\
        L = -\tau_7^3 J_{max,i}+\tau_1^3 J_{max,i}+2 \tau_1 V_{In}+2 \tau_7 V_F'
      \end{array}
    \right.
  \end{equation}

\begin{figure}[!ht]
  \centering
  \caption{Case 6\label{fig:case6}}    
  \includegraphics[width=0.4\textwidth]{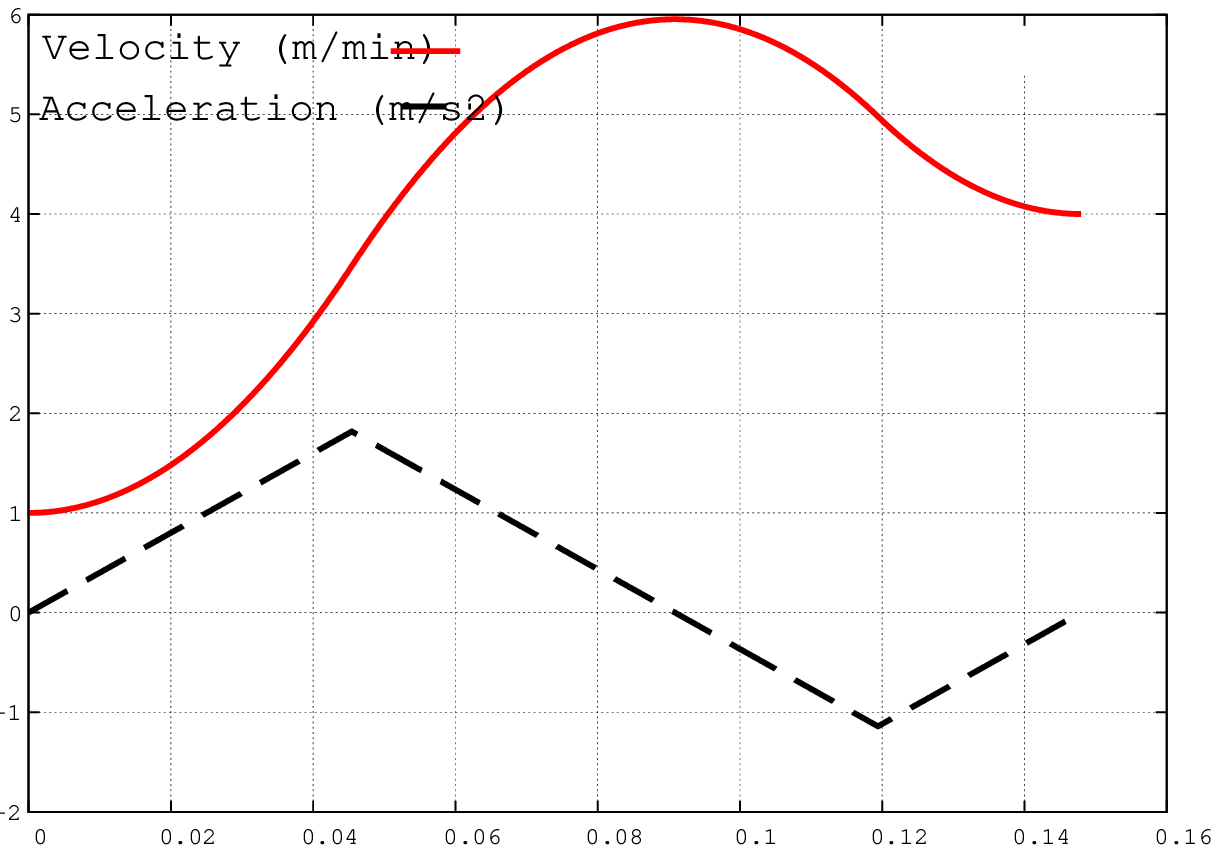}\\
\end{figure}

\item Case 7: $V_{Out}$ is reached by reaching $A_{max,i}$. $V_F$
  is reached. $V_F$ is reached by reaching $A_{max,i}$. 
  \begin{equation}
    \left\{
      \begin{array}{l}
        |V_{F} - V_{In} | \geq \frac{A_{max,i}^2}{J_{max,i}}\\
        |V_{Out} - V_{F} | \geq \frac{A_{max,i}^2}{J_{max,i}}\\
        \begin{array}{rcl}
          L &\geq &\frac{1}{2 A_{max,i} J_{max,i}} \cdot \\
        && \Big(
        \left(2 J_{max,i} V_F+2 A_{max,i}^2\right) V_{In}\\
        && + J_{max,i} V_F^2+A_{max,i}^2V_F \\ 
        && + \left(2 J_{max,i} V_{Out}+2 A_{max,i}^2\right) V_{F}\\
        &&+J_{max,i} V_{Out}^2+A_{max,i}^2V_{Out} \Big) \\ 
        \end{array}
      \end{array}
    \right.
    \label{eq:7.cond}
  \end{equation}
  \begin{equation}
    \left\{
      \begin{array}{c}
        \tau_1 = \tau_3 = \tau_5 = \tau_7 =  \frac{A_{max}}{J_{max}}\\
        \tau_2 = \frac{V_{F}}{A_{max}} - \frac{A_{max}}{J_{max}} \\
        \tau_6 = \frac{V_{Out}}{A_{max}} - \frac{A_{max}}{J_{max}}
      \end{array}
    \right.
    \label{eq:7.res}
  \end{equation}
\item Case 8: $V_{Out}$ is reached without reaching $A_{max,i}$. $V_F$
  is reached. $V_F$ is reached by reaching $A_{max,i}$. 
 \begin{equation}
    \left\{
      \begin{array}{l}
        |V_{F} - V_{In} | \geq \frac{A_{max,i}^2}{J_{max,i}}\\
        |V_{Out} - V_{F} | \leq \frac{A_{max,i}^2}{J_{max,i}}\\
        \begin{array}{rcl}
          L &\geq &\frac{\left(2 J_{max,i} V_F+2 A_{max,i}^2\right)
            V_{In}+J_{max,i} V_F^2+A_{max,i}^2V_F}{2 A_{max,i}
            J_{max,i}} \\
          & + & \frac{\sqrt{J_{max}V_{Out}}\left(4V_{F}+V_{Out} \right)+A_{max}
            V_F}{2 J_{max}} \\ 
        \end{array}
      \end{array}
    \right.  
    \label{eq:8.cond}
  \end{equation}
  \begin{equation}
    \left\{
      \begin{array}{l}
        \tau_1 = \tau_3 = \frac{A_{max,i}}{J_{max,i}} \\
        \tau_2 = \frac{V_F-V_{In}}{A_{max,i}}-\frac{A_{max,i}}{J_{max,i}}\\
        \tau_5 = \tau_7  = \sqrt{\frac{V_F - V_{Out}}{J_{max,i}}}\\
        \tau_6 = 0
      \end{array}
    \right.  
    \label{eq:8.res}
  \end{equation}

\item Case 9: $V_{Out}$ is reached by reaching $A_{max,i}$. $V_F$
  is reached. $V_F$ is reached without reaching $A_{max,i}$. 
  \begin{equation}
    \left\{
      \begin{array}{l}
        |V_{F} - V_{In} | \leq \frac{A_{max,i}^2}{J_{max,i}} \\
        |V_{Out} - V_{F} | \geq \frac{A_{max,i}^2}{J_{max,i}} \\
        \begin{array}{rcl}
          L &\geq & \frac{\sqrt{J_{max}V_{F}}\left(4V_{In}+V_{F} \right)+A_{max}
            V_{In}}{2 J_{max}} \\
          & + &\frac{\left(2 J_{max,i} V_{Out}+2 A_{max,i}^2\right)
            V_{F}+J_{max,i} V_{Out}^2+A_{max,i}^2V_{Out}}
          {2 A_{max,i} J_{max,i}}
        \end{array}
      \end{array}    
    \right.
    \label{eq:9.cond}
  \end{equation} 
  \begin{equation}
    \left\{
      \begin{array}{l}
        \tau_1 = \tau_3 = \sqrt{\frac{V_{In} - V_{F}}{J_{max,i}}}\\
        \tau_2 =0 \\
        \tau_5 = \tau_7  = \frac{A_{max,i}}{J_{max,i}}\\
        \tau_6 =  \frac{V_{Out}-V_{F}}{A_{max,i}}-\frac{A_{max,i}}{J_{max,i}}
      \end{array}
    \right.  
    \label{eq:9.res}
  \end{equation}

\item Case 10: $V_{Out}$ is reached without reaching $A_{max,i}$. $V_F$
  is reached. $V_F$ is reached without reaching $A_{max,i}$ (figure \ref{fig:case10}) :  
  \begin{equation}
    \left\{
      \begin{array}{l}
        |V_{F} - V_{In} | \leq \frac{A_{max,i}^2}{J_{max,i}} \\
        |V_{Out} - V_{F} | \leq  \frac{A_{max,i}^2}{J_{max,i}} \\   
        \begin{array}{rcl}
          L &\geq & \frac{\sqrt{J_{max}V_{F}}\left(4V_{In}+V_{F} \right)+A_{max}
            V_{In}}{2 J_{max}}  \\
          &+&  \frac{\sqrt{J_{max}V_{Out}}\left(4V_{F}+V_{Out} \right)+A_{max}
            V_F}{2 J_{max}} \\ 
        \end{array}
      \end{array}
    \right.
    \label{eq:10.cond}
  \end{equation}
  \begin{equation}
    \left\{
      \begin{array}{l}
        \tau_1 = \tau_3 = \sqrt{\frac{V_{In} - V_{F}}{J_{max,i}}}\\
        \tau_2 =0 \\
        \tau_5 = \tau_7  = \sqrt{\frac{V_{F} - V_{Out}}{J_{max,i}}}\\
        \tau_6 =  0
      \end{array}
    \right.  
    \label{eq:10.res}
  \end{equation}
\end{itemize}

\begin{figure}[!ht]
  \centering
  \caption{Case 10\label{fig:case10}}    
  \includegraphics[width=0.4\textwidth]{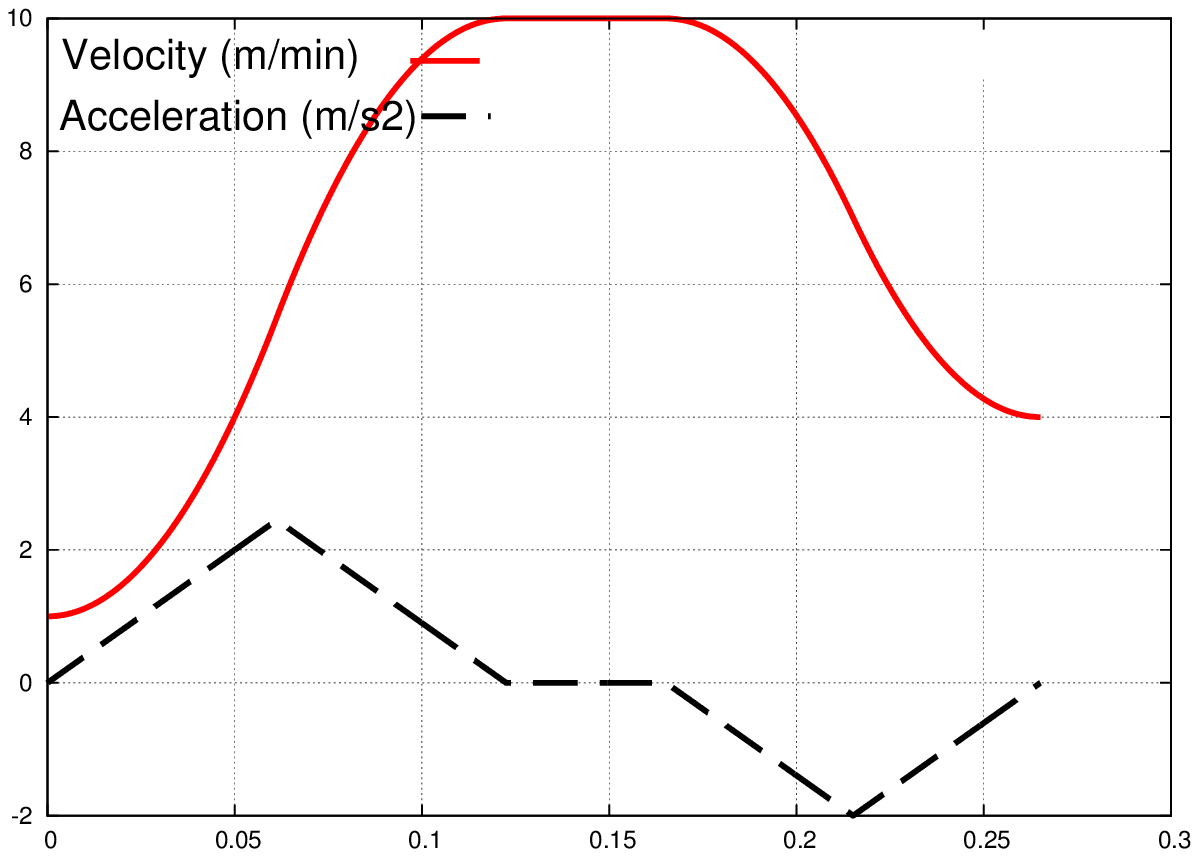}
\end{figure}

In the simulations, equation 2 is resolved by the Cardan method. The systems of equation \ref{eq:3.res}, \ref{eq:4.res},
\ref{eq:5.res} and \ref{eq:6.res} are non-linear. They are resolved numerically using the Newton-Raphson method.

To use this algorithm, take a segment of $0.01 m$ to be covered. Take $V_{In} = 0.2 m/s$, $V_{Out} = 0.1 m/s$, $V_{F} = 0.5
m/s$. The machine parameters are those in table \ref{tab:caracteristiques}. 

$|V_{Out} - V_{In}| \leq \frac{A_{max,i}^2}{J_{max,i}}$; thus there will be no phase 2. $L \geq \left(V_{Out} +V_{In} \right)
\sqrt{\frac{V_{Out} -V_{In}}{J_{max}}}$; the length of the segment thus allows $V_{Out}$ to be reached. 

$|V_{F} - V_{In}| \leq \frac{A_{max,i}^2}{J_{max,i}}$ and $|V_{Out} -
V_{F}| \leq \frac{A_{max,i}^2}{J_{max,i}}$ and $L \leq
\frac{\sqrt{J_{max}V_{Out}}\left(4V_{F}+V_{Out} \right)+A_{max} V_F}{2
  J_{max}}$. The length does therefore not allow $V_F$ to be reached and the differences in feed rates will not allow maximum acceleration to be reached. This means case 6 applies. This gives$V_F' = 0.4418 m/s$, 
$$
    \tau_1 = \tau_3 = 0.07776s. \quad   \tau_5 = \tau_7 = 0.09245s.
$$

\end{document}